\documentclass{lmcs}
\pdfoutput=1
\usepackage[utf8]{inputenc}

\usepackage{lastpage}
\lmcsdoi{21}{4}{20}
\lmcsheading{}{\pageref{LastPage}}{}{}%
{Mar.~13,~2024}{Nov.~07,~2025}{}

\usepackage{lineno}
\usepackage{enumitem}
\usepackage{graphicx}	
\usepackage{amsmath, amsthm, amssymb}
\usepackage{xspace}
\usepackage{comment}
\usepackage{hyperref}
\usepackage{xcolor}
\newcommand{\algprobm}[1]{\textsc{#1}\xspace}

\newcommand{\Soc}{\text{Soc}}
\newcommand{\Fac}{\text{Fac}}

\theoremstyle{definition}
\newtheorem{question}[thm]{Question}

\newcommand{\Lem}[1]{Lem.~\ref{#1}\xspace}

\newcommand{\Prop}[1]{Prop.~\ref{#1}\xspace}
\newcommand{\Thm}[1]{Thm.~\ref{#1}\xspace}

\newcommand{\wt}{\text{wt}}

\DeclareMathOperator{\Aut}{Aut}

\DeclareMathOperator{\rad}{Rad}
\DeclareMathOperator{\pker}{PKer}

\begin{document}
\title[Descriptive Complexity of Semisimple Groups]{On the Descriptive Complexity of Groups\texorpdfstring{\\}{} without Abelian Normal Subgroups}

\author[J.A.~Grochow]{Joshua A. Grochow\lmcsorcid{0000-0002-6466-0476}}[a,b]
\address{Department of Computer Science, University of Colorado Boulder, 1111 Engineering Dr, ECOT 717, 430 UCB, Boulder, CO 80309, USA}
\email{jgrochow@coolorado.edu}

\address{Department of Mathematics, University of Colorado Boulder, Campus Box 395, Boulder, CO 80309 USA}

\author[M.~Levet]{Michael Levet\lmcsorcid{0009-0009-1992-3175}}[c]
\address{Department of Computer
Science, College of Charleston, 66
George Street, Charleston, SC
29424, USA}
\email{levetm@cofc.edu}

\begin{abstract}
In this paper, we explore the descriptive complexity theory of finite groups by examining the power of the second Ehrenfeucht--Fraïss\'e bijective pebble game in Hella's (\textit{Ann. Pure Appl. Log.}, 1989) hierarchy. This is a Spoiler--Duplicator game in which Spoiler can place up to two pebbles each round. While it trivially solves graph isomorphism, it may be nontrivial for finite groups, and other ternary relational structures. We first provide a novel generalization of Weisfeiler--Leman (WL) coloring, which we call \textit{2-ary} WL. We then show that 2-ary WL is equivalent to the second Ehrenfeucht--Fra\"iss\'e bijective pebble game in Hella's hierarchy. 

Our main result is that, in the pebble game characterization, only $O(1)$ pebbles and $O(1)$ rounds are sufficient to identify all groups without Abelian normal subgroups (a class of groups for which isomorphism testing is known to be in $\mathsf{P}$; 
Babai, Codenotti, \& Qiao, ICALP 2012). We actually show that $7$ pebbles and $7$ rounds suffice. In particular, we show that within the first few rounds, Spoiler can force Duplicator to select an isomorphism between two such groups at each subsequent round. By Hella's results (\emph{ibid.}), this is equivalent to saying that these groups are identified by formulas in first-order logic with generalized 2-ary quantifiers, using only $O(1)$ variables and $O(1)$ quantifier depth. 
\end{abstract}

\maketitle

\section{Introduction} \label{sec:introduction}

\footnote{A preliminary version of this work previously appeared in GandALF 2023 \cite{GLGandalf2023}. Here, we provide full proofs of (i) the equivalence between $2$-ary WL and Hella's $2$-ary pebble game, and (ii) key lemmas for establishing our main result Theorem~\ref{thm:SemisimpleMain}. Due to space constraints, these details were omitted in the extended abstract that appeared in GandALF.
} Descriptive complexity theory studies the relationship between the complexity of describing a given problem in some logic, and the complexity of solving the problem by an algorithm. When the problems involved are isomorphism problems, Immerman and Lander \cite{ImmermanLander1990} showed that complexity of a logical sentence describing the isomorphism type of a graph was essentially the same as the Weisfeiler--Leman coloring dimension of that graph, and the complexity of an Ehrenfeucht--Fra\"iss\'e pebble game (see also \cite{CFI}). 

It is a well-known open question whether there is a logic that exactly captures the complexity class $\textsf{P}$ on unordered (unlabeled) structures; on ordered structures such a logic was given by Immerman \cite{ImmermanPTime} and Vardi \cite{VardiPTime}. The difference between these two settings is essentially the \textsc{Graph Canonization} problem, whose solution allows one to turn an unordered graph into an ordered graph in an isomorphism-preserving way.

One natural approach in trying to capture $\textsf{P}$ on unordered structures is thus to attempt to extend first-order logic $\textsf{FO}$ by generalized quantifiers (cf. Mostowski \cite{Mostowski1957} and Lindstrom \cite{Lindstrom}) in the hopes that the augmented logics can characterize finite graphs up to isomorphism, thus reducing the unordered case to the previously solved ordered case. A now-classical approach, initiated by Immerman \cite{ImmermanPTime}, was to augment fixed-point logic with counting quantifiers, which can be analyzed in terms of an equivalence induced by (variable confined) fragments of first-order logic with counting. However, Cai, F\"urer, \& Immerman \cite{CFI} showed that $\textsf{FO}+\textsf{LFP}$ plus counting does not capture $\textsf{P}$ on finite graphs. More generally, Flum \& Grohe have characterized when $\textsf{FO}$ plus counting captures $\textsf{P}$ on unordered structures \cite{GroheFlum}.

The approach of Cai, F\"urer, \& Immerman (\emph{ibid.}, see also \cite{ImmermanLander1990}) was to prove a three-way equivalence: between (1) counting logics, (2) the higher-dimensional Weisfeiler--Leman coloring procedure, and (3) Ehrenfeucht--Fra\"iss\'e pebble games. Ehrenfeucht--Fra\"iss\'e pebble games \cite{Ehrenfeucht,Fraisse} have long been an important tool in proving the inexpressibility of certain properties in various logics; in this case, they used such games to show that the logics could not express the difference between certain pairs of non-isomorphic graphs. Consequently, Cai, F\"urer, \& Immerman ruled out the Weisfeiler--Leman algorithm as a polynomial-time isomorphism test for graphs, which resolved a long-standing open question in isomorphism testing. Nonetheless, the Weisfeiler--Leman coloring procedure is a key subroutine in many algorithms for \textsc{Graph Isomorphism}, including Babai's quasi-polynomial-time algorithm \cite{BabaiGraphIso}. It is thus interesting to study its properties and its distinguishing power.

While the result of Cai, F\"urer, \& Immerman ruled out Weisfeiler--Leman as a polynomial-time isomorphism test for graphs, for \emph{groups} it remains an interesting open question. The general WL procedure for groups was introduced by Brachter \& Schweitzer \cite{WLGroups} and has been studied in several papers since then \cite{BrachterSchweitzerWLLibrary,GrochowLevetWL,CollinsLevetWL,CollinsUndergradThesis,LevetThesis,VagnozziThesis,BrachterThesis,ChenRenPonomarenko}. Outside the scope of WL, it is known that \textsc{Group Isomorphism} is $\textsf{AC}^{0}$-reducible to \textsc{Graph Isomorphism}, and there is no $\textsf{AC}^0$ reduction in the opposite direction \cite{ChattopadhyayToranWagner}. For this and other reasons, \textsc{Group Isomorphism} is believed to be the easier of the two problems, so it is possible that WL---and more generally, tools from descriptive complexity---could yield stronger results for groups than for graphs.

On graphs, which are binary relational structures, if Spoiler is allowed to pebble two elements per turn, then Spoiler can win on any pair of non-isomorphic graphs. However, groups are ternary relational structures (the relation is $\{(a,b,c) : ab=c\}$), so such a game may yield nontrivial insights into the descriptive complexity of finite groups. Hella \cite{Hella1989, Hella1993} introduced such games in a more general context, and showed that allowing Spoiler to pebble $q$ elements per round corresponded to the generalized $q$-ary quantifiers of Mostowski \cite{Mostowski1957} and Lindstrom \cite{Lindstrom}. When $q=1$, Hella shows that this pebble game is equivalent in power to the $\textsf{FO}$ plus counting logics mentioned above. Our focus in this paper is to study the power of the $q=2$-ary game for identifying finite groups. 

\noindent \\ \textbf{Main Results.} In this paper, we initiate the study of Hella's $2$-ary Ehrenfeucht--Fra\"iss\'e-style pebble game, in the setting of groups. Our main result is that this pebble game efficiently characterizes isomorphism in a class of groups for which isomorphism testing is known to be in $\mathsf{P}$, but only by quite a nontrivial algorithm, which we discuss after stating our result. 

\begin{thm} \label{thm:SemisimpleMain}
Let $G$ be a group with no Abelian normal subgroups (a.k.a. Fitting-free or semisimple), and let $H$ be arbitrary. If $G \not \cong H$, then Spoiler has a winning strategy for the Ehrenfeucht--Fra\"iss\'e game at the second level of Hella's hierarchy, using $7$ pebbles and $7$ rounds.
\end{thm}

Every group $G$ can be written as an extension of its solvable radical $\rad(G)$ by the quotient $G/\rad(G)$, which does not have Abelian normal subgroups. As such, the latter class of groups is quite natural, both group-theoretically and computationally. Computationally, it has been used in algorithms for general finite groups both in theory (e.g., \cite{Babai1999GroupsSA,BabaiBealsSeress}) and in practice (e.g., \cite{CH03}). Isomorphism testing in this family of groups can be solved efficiently in practice \cite{CH03}, and is known to be in $\mathsf{P}$ through a series of two papers \cite{BCGQ, BCQ}. 

In proving \Thm{thm:SemisimpleMain}, we show that with the use of only a few (at most $7$) pebbles, Spoiler can effectively force Duplicator to select an isomorphism of $G$ and $H$. We contrast this with the setting of ordinary ($1$-ary) Weisfeiler--Leman (which is equivalent to the $1$-ary pebble game): subsequent to the preprint of our paper \cite{GLGandalf2023}, Brachter improved the current-best upper bound on the classical $1$-ary WL-dimension from the trivial $\log n$ down to $O(\log \log n)$ \cite{BrachterThesis}. Furthermore, we do not have any lower bounds on the WL-dimension for semisimple groups.

\noindent \\ \textbf{Further Related Work.} Despite the fact that Weisfeiler--Leman is insufficient to place \algprobm{Graph Isomorphism} (\algprobm{GI}) into $\textsf{PTIME}$, it remains an active area of research. For instance, Weisfeiler--Leman is a key subroutine in Babai's quasipolynomial-time \algprobm{GI} algorithm \cite{BabaiGraphIso}. Furthermore, Weisfeiler--Leman has led to advances in simultaneously developing both efficient isomorphism tests and the descriptive complexity theory for finite graphs; see for instance, \cite{GroheBook, GroheVerbitsky, KieferMcKay, KieferPonomarenkoSchweitzer, grohe2019linear, grohe2019rankwidth, grohe2021logarithmic, KieferSchweitzerSelman, ajtai_fagin_1990, ARORA199797, Rossman2009EhrenfeuchtFrassGO}. Weisfeiler--Leman also has close connections to the Sherali--Adams hierarchy in linear programming  \cite{GroheOttoLinearEquations}.

The complexity of the \algprobm{Group Isomorphism} (\algprobm{GpI}) problem is a well-known open question. In the Cayley (multiplication) table model, $\algprobm{GpI}$ belongs to $\textsf{NP} \cap \textsf{coAM}$. The generator-enumerator algorithm, attributed to Tarjan in 1978 \cite{MillerTarjan}, has time complexity $n^{\log_{p}(n) + O(1)}$, where $n$ is the order of the group and $p$ is the smallest prime dividing $n$. This bound has escaped largely unscathed: Rosenbaum \cite{Rosenbaum2013BidirectionalCD} (see \cite[Sec. 2.2]{GR16}) improved this to $n^{(1/4)\log_p(n) + O(1)}$. And even the impressive body of work on practical algorithms for this problem, led by Eick, Holt, Leedham-Green and O'Brien (e.\,g., \cite{BEO02, ELGO02, BE99, CH03}) still results in an $n^{\Theta(\log n)}$-time algorithm in the general case (see \cite[Page 2]{WilsonSubgroupProfiles}). In the past several years, there have been significant advances on algorithms with worst-case guarantees on the serial runtime for special cases of this problem including Abelian groups \cite{Kavitha, Vikas, Savage}, direct product decompositions \cite{WilsonDirectProductsArxiv, KayalNezhmetdinov}, groups with no Abelian normal subgroups \cite{BCGQ, BCQ}, coprime and tame group extensions \cite{Gal09,QST11,BQ, GQ15}, low-genus $p$-groups and their quotients \cite{LW12,BMWGenus2}, Hamiltonian groups \cite{DasSharma}, and groups of almost all orders \cite{DietrichWilson}.

Key motivation for $\algprobm{GpI}$ is due to its close relation to $\algprobm{GI}$. In the Cayley (verbose) model, $\algprobm{GpI}$ reduces to $\algprobm{GI}$ \cite{ZKT}, 
while $\algprobm{GI}$ reduces to the succinct $\algprobm{GpI}$ problem \cite{Heineken1974TheOO, Mekler} (recently simplified \cite{HeQiao}). In light of Babai's breakthrough result that $\algprobm{GI}$ is quasipolynomial-time solvable \cite{BabaiGraphIso}, $\algprobm{GpI}$ in the Cayley model is a key barrier to improving the complexity of $\algprobm{GI}$. Both verbose $\algprobm{GpI}$ and $\algprobm{GI}$ are considered to be candidate $\textsf{NP}$-intermediate problems, that is, problems that belong to $\textsf{NP}$, but are neither in $\textsf{P}$ nor $\textsf{NP}$-complete \cite{Ladner}. There is considerable evidence suggesting that $\algprobm{GI}$ is not $\textsf{NP}$-complete \cite{Schoning, BuhrmanHomer, ETH, BabaiGraphIso, GILowPP, ArvindKurur}. As verbose $\algprobm{GpI}$ reduces to $\algprobm{GI}$, this evidence also suggests that $\algprobm{GpI}$ is not $\textsf{NP}$-complete. It is also known that $\algprobm{GI}$ is strictly harder than $\algprobm{GpI}$ under $\textsf{AC}^{0}$ reductions \cite{ChattopadhyayToranWagner}. 

While the descriptive complexity of graphs has been extensively studied, the work on the descriptive complexity of groups is scant compared to the algorithmic literature on \algprobm{Group Isomorphism} (\algprobm{GpI}). There has been work relating first order logics and groups \cite{FiniteGroupsFOL}, as well as work examining the descriptive complexity of finite Abelian groups \cite{DescriptiveComplexityAbelianGroups}. Recently, Brachter \& Schweitzer \cite{WLGroups} introduced three variants of Weisfeiler--Leman for groups, including corresponding logics and pebble games. These pebble games correspond to the first level of Hella's hierarchy \cite{Hella1989, Hella1993}. In particular, Brachter \& Schweitzer showed that $3$-dimensional Weisfeiler--Leman can distinguish $p$-groups arising from the CFI graphs \cite{CFI} via Mekler's construction \cite{Mekler}, suggesting that $\textsf{FO} + \textsf{LFP} + \textsf{C}$ may indeed capture $\textsf{PTIME}$ on groups. Determining whether even $o(\log n)$-dimensional Weisfeiler--Leman can resolve \algprobm{GpI} is an open question.

The use of Weisfeiler--Leman for groups is quite new. To the best of our knowledge, using Weisfeiler--Leman for \algprobm{Group Isomorphism} testing was first attempted by Brooksbank, Grochow, Li, Qiao, \& Wilson \cite{BGLQW}. Brachter \& Schweitzer \cite{WLGroups} subsequently introduced three variants of Weisfeiler--Leman for groups that more closely resemble that of graphs. In particular, Brachter \& Schweitzer \cite{WLGroups}  characterized their algorithms in terms of logics and Ehrenfeucht--Fra\"iss\'e pebble games. The relationship between the works of Brachter \& Schweitzer and Brooksbank, Grochow, Li, Qiao, \& Wilson \cite{BGLQW} is an interesting question.

In subsequent work, Brachter \& Schweitzer \cite{BrachterSchweitzerWLLibrary} further developed the descriptive complexity of finite groups. They showed in particular that low-dimensional Weisfeiler--Leman can detect key group-theoretic invariants such as composition series, radicals, and quotient structure. Furthermore, they also showed that Weisfeiler--Leman can identify direct products in polynomial-time, provided it can also identify the indecomposable direct factors in polynomial-time. Grochow \& Levet \cite{GrochowLevetWL} extended this result to show that Weisfeiler--Leman can compute direct products in parallel, provided it can identify each of the indecomposable direct factors in parallel. Additionally, Grochow \& Levet showed that constant-dimensional Weisfeiler--Leman can in a constant number of rounds identify coprime extensions $H \ltimes N$, where the normal Hall subgroup $N$ is Abelian and the complement $H$ is $O(1)$-generated. This placed isomorphism testing into $\textsf{L}$; the previous bound for isomorphism testing in this family was $\textsf{P}$ \cite{QST11}. Grochow \& Levet also ruled out $\textsf{FO} + \textsf{LFP}$ as a candidate logic for capturing $\textsf{PTIME}$ on finite groups, by showing that the count-free Weisfeiler--Leman algorithm cannot even identify Abelian groups in polynomial-time.

\section{Preliminaries}

\subsection{Groups}

Unless stated otherwise, all groups are assumed to be finite. For a set $X$, let $\text{Sym}(X)$ be the group of permutations on $X$. The \textit{socle} of a group $G$, denoted $\text{Soc}(G)$, is the subgroup generated by the minimal normal subgroups of $G$. If $G$ has no Abelian normal subgroups, then $\text{Soc}(G)$ decomposes uniquely as the direct product of non-Abelian simple factors. The Classification of Finite Simple Groups provides that, if $S$ is a non-Abelian simple group, then there exist $x, y \in S$ such that $S = \langle x, y \rangle$, and this is the only way in which our results depend on the Classification. The \textit{normal closure} of a subset $S \subseteq G$, denoted $\text{ncl}(S)$, is the smallest normal subgroup of $G$ that contains $S$. 

We say that a normal subgroup $N \trianglelefteq G$ \textit{splits} in $G$ if there exists a subgroup $H \leq G$ such that $H \cap N = \{1\}$ and $G = HN$. The conjugation action of $H$ on $N$ allows us to express multiplication of $G$ in terms of pairs $(h, n) \in H \times N$. We note that the conjugation action of $H$ on $N$ induces a group homomorphism $\theta : H \to \Aut(N)$ mapping $h \mapsto \theta_{h}$, where $\theta_{h} : N \to N$ sends $\theta_{h}(n) = hnh^{-1}$. So given $(H, N, \theta)$, we may define the group $H \ltimes_{\theta} N$ on the set $\{ (h, n) : h \in H, n \in N \}$ with the product $(h_{1}, n_{1})(h_{2}, n_{2}) = (h_{1}h_{2}, \theta_{h_{2}^{-1}}(n_{1})n_{2})$. We refer to the decomposition $G = H \ltimes_{\theta} N$ as a \textit{semidirect product} decomposition. When the action $\theta$ is understood, we simply write $G = H \ltimes N$. If $\theta$ is trivial, then $H \trianglelefteq G$ and $G = H \times N$. In this case, we say that $H$ is a \textit{direct complement} of $N$ in $G$. The \textit{wreath product} $N \wr \text{Sym}(k)$ is the semidirect product $\text{Sym}(k) \ltimes N^k$, where $\text{Sym}(k)$ acts by permuting the copies of $N$.

For $g, h \in G$, the \textit{commutator} $[g, h] := ghg^{-1}h^{-1}$. If $X, Y \leq G$, then $[X, Y] = \langle \{ [x, y] : x \in X, y \in Y \} \rangle.$

\subsection{Pebbling Game}

We recall the bijective pebble game of Hella \cite{Hella1989, Hella1993}, in the context of WL on graphs as that is likely more familiar to more readers. This game is often used to show that two graphs $X$ and $Y$ cannot be distinguished by $k$-WL. The game is an Ehrenfeucht--Fra\"iss\'e game, with two players: Spoiler and Duplicator. Each player begins with $k+1$ pebbles, $p_1, \dotsc, p_{k+1}$ for Spoiler and $p_1^{\prime}, \dotsc, p_{k+1}^{\prime}$ for Duplicator, which are placed beside the graph. Each round proceeds as follows.
\begin{enumerate}
\item Spoiler chooses $i \in [k+1]$, and picks up pebbles $p_i, p^{\prime}_i$.
\item Duplicator chooses a bijection $f : V(X) \to V(Y)$.
\item Spoiler places $p_{i}$ on some vertex $v \in V(X)$. Then $p_{i}^{\prime}$ is placed on $f(v)$. 
\item We check the winning condition, which will be formalized later.
\end{enumerate} 

Let $v_{1}, \ldots, v_{m}$ be the vertices of $X$ pebbled at the end of step 1, and let $v_{1}^{\prime}, \ldots, v_{m}^{\prime}$ be the corresponding pebbled vertices of $Y$. Spoiler wins precisely if the map $v_{\ell} \mapsto v_{\ell}^{\prime}$ does not extend to an isomorphism of the induced subgraphs $X[\{v_{1}, \ldots, v_{m}\}]$ and $Y[\{v_{1}^{\prime}, \ldots, v_{m}^{\prime}\}]$. Duplicator wins otherwise. There is one additional rule: if $X$ and $Y$ do not have the same number of vertices, then we say that Spoiler wins before round 1 begins, or ``after 0 rounds.''  
We note that $X$ and $Y$ are not distinguished by the first $r$ rounds of $k$-WL if and only if Duplicator wins the first $r$ rounds of the $(k+1)$-pebble game \cite{Hella1989, Hella1993, CFI}.

Hella \cite{Hella1989, Hella1993} exhibited a hierarchy of ($q$-ary) pebble games where, for $q \geq 1$, Spoiler could pebble a sequence of $1 \leq j \leq q$ elements $(v_{1}, \ldots, v_{j}) \mapsto (f(v_{1}), \ldots, f(v_{j}))$ in a single round. The case of $q = 1$ corresponds to the case of Weisfeiler--Leman. As remarked by Hella \cite[p.~6, just before \S 4]{Hella1993}, the $q$-ary game immediately identifies all relational structures of arity $\leq q$. For example, the $q=2$ game on graphs solves $\algprobm{GI}$: for if two graphs $X$ and $Y$ are non-isomorphic, then any bijection $f : V(X) \to V(Y)$ that Duplicator selects must map an adjacent pair of vertices $u,v$ in $X$ to a non-adjacent pair $f(u), f(v)$ in $Y$ or vice-versa. Spoiler immediately wins by pebbling $(u, v) \mapsto (f(u), f(v))$. However, as groups are ternary relational structures (the relation being $\{(a,b,c) : a,b,c \in G, ab=c\}$), the $q=2$ case can, at least in principle, be non-trivial on groups. 

Brachter \& Schweitzer \cite{WLGroups} adapted Hella's \cite{Hella1989, Hella1993} pebble games in the $q = 1$ case to the setting of groups, obtaining three different versions. Their Version III involves reducing to graphs and playing the pebble game on graphs, so we don't consider it further here. Versions I and II are both played on the groups $G$ and $H$ directly.

Both versions are played identically as for graphs, with the only difference being the winning condition. We recall the following standard definitions in order to describe these winning conditions.

\begin{defi}
Let $G,H$ be two groups. Given $k$-tuples $\overline{g} = (g_1, \dotsc, g_k) \in G^k$ and $\overline{h} = (h_1, \dotsc, h_k) \in H^k$, we say $(\overline{g}, \overline{h})$ ...
\begin{enumerate}
\item ...\emph{gives a well-defined map} if $g_i = g_j \Leftrightarrow h_i = h_j$ for all $i \neq j$;

\item ...are \emph{partially isomorphic} or \emph{give a partial isomorphism} if they give a well-defined map, and for all $i,j,k$ we have $g_i g_j = g_k \Leftrightarrow h_i h_j = h_k$;

\item ...are \emph{marked isomorphic} or \emph{give a marked isomorphism} if it gives a well-defined map, and the map that sends $g_i \mapsto h_i$ for all $i$ extends to an isomorphism $\langle g_1, \dotsc, g_k \rangle \to \langle h_1, \dotsc, h_k \rangle$.
\end{enumerate}
\end{defi}

Let $v_{1}, \ldots, v_{m}$ be the group elements of $G$ pebbled at the end of step 1, and let $v_{1}^{\prime}, \ldots, v_{m}^{\prime}$ be the corresponding pebbled vertices of $H$. In Version I, Spoiler wins precisely if $(\overline{v}, \overline{v}^{\prime})$ does not give a partial isomorphism, and in Version II Spoiler wins precisely if $(\overline{v}, \overline{v}^{\prime})$ does not give a marked isomorphism.

\textit{Notation:} If $f$ is the bijection chosen by Duplicator, and Spoiler pebbles $g \in G$, resp. the pair $g_1,g_2 \in G$, we write ``Spoiler pebbles $g \mapsto f(g)$'', resp. ``Spoiler pebbles $(g_1,g_2) \mapsto (f(g_1), f(g_2))$''.

Both Versions I and II may be generalized to allow Spoiler to pebble up to $q$ group elements at a single round, for some $q \geq 1$. Mimicking the proof above for $q=2$ for graphs, we have that $q = 3$ is sufficient to solve $\algprobm{GpI}$ in a single round. The distinguishing power, however, of the $q = 2$ game for groups remains unclear, and is the main subject of this paper. As we are interested in the round complexity, we introduce the following notation. 

\begin{defi}[Notation for pebbles, rounds, arity, and WL version]
Let $k \geq 2, r \geq 1$, $q \in [3]$, and $J \in \{I, II\}$. Denote $(k,r)$-WL$^q_J$ to be the $k$-pebble, $r$-round, $q$-ary Version $J$ pebble game.
\end{defi}

We refer to $q$ as the \emph{arity} of the pebble game, as it corresponds to the arity of generalized quantifiers in a logic whose distinguishing power is equivalent to that of the game:

\begin{rem}[Equivalence with logics with generalized $2$-ary quantifiers]
Hella \cite{Hella1989} describes the game (essentially the same as our description, but with no restriction on number of pebbles, and a transfinite number of rounds) for general $q$ at the bottom of p.~244, for arbitrary relational structures. We restrict to the case of $q = 2$, a finite number of pebbles and rounds, and the (relational) language of groups. Hella proves that this game is equivalent to first-order logic with arbitrary $q$-ary equantifiers in \cite[Thm.~2.5]{Hella1989}.
\end{rem}

We will frequently use the following observation without mention.

\begin{obs} \label{obs:inverse}
In the $2$-ary Version II pebble game, Duplicator must select a bijection that preserves the identity, or Spoiler wins in that round. If Spoiler has two pebbles in hand, Duplicator must select a bijection that preserves inverses, or Spoiler wins in that round.
\end{obs}

\begin{proof}
Suppose not. If Duplicator selects a bijection that does not preserve the identity element, then Spoiler places a pebble on $1 \mapsto f(1) \neq 1$, and wins at the end of that round as $1 \cdot 1 = 1$, but $f(1) \cdot f(1) \neq f(1)$.

Next, suppose Spoiler has two pebbles in hand, and Duplicator selects a bijection $f : G \to H$ such that $f(g^{-1}) \neq f(g)^{-1}$. Spoiler pebbles $(g, g^{-1}) \mapsto (f(g), f(g^{-1}))$. Now $gg^{-1} = 1$, while $f(g)f(g^{-1}) \neq 1$. So Spoiler wins at the end of this round.
\end{proof}

\section{Higher-arity Weisfeiler-Leman-style coloring\texorpdfstring{\\}{} corresponding to higher arity pebble games} \label{sec:coloring}
Brachter \& Schweitzer \cite{WLGroups} introduced Versions I and II of $1$-ary WL, which are equivalent up to a small additive constant in the WL-dimension \cite{WLGroups} and $O(\log n)$ rounds \cite{GrochowLevetWL}. For the purpose of comparison, we introduce Versions I and II of $2$-ary WL. We will see later that only one additional round suffices in the $2$-ary case (see \Thm{thm:rounds}). The differences in Versions I and II of WL (both the $1$-ary and $2$-ary variants) arise from whether the group is viewed as a structure with a ternary relation (Version I) or as a structure with a binary function (Version II).

To give our definition of 2-ary color refinement, we introduce some notation and terminology. Given a $k$-tuple $\overline{x} = (x_1, \dotsc, x_k) \in G^k$, a pair of distinct indices $i,j \in [k]$, and a pair of group elements $y, z$, we define $\overline{x}_{(i,j) \leftarrow (y,z)}$ to be the $k$-tuple $\overline{x}'$ that agrees with $\overline{x}$ on all indices besides $i,j$, and with $x_i' = y, x_j' = z$. If $i=j$, we require $y=z$, and we denote this $\overline{x}_{i \leftarrow y}$. Two graphs $\Gamma_1, \Gamma_2$, with edge-colorings $c_i \colon E(\Gamma_i) \to C$ to some color set $C$ (for $i=1,2$) are color isomorphic if there is a graph isomorphism $\varphi \colon V(\Gamma_1) \to V(\Gamma_2)$ that also preserves colors, in the sense that $c_1((u,v)) = c_2((\varphi(u), \varphi(v))$ for all edges $(u,v) \in E(\Gamma_1)$. We note that the color names for the group elements must be consistent with the color names for the complete graphs in the refinement steps. Beyond this consistency, the names of the colors themselves are not relevant.

\begin{defi}[2-ary $k$-dimensional Weisfeiler-Leman coloring]
Let $G,H$ be two groups of the same order, let $k \geq 1$.
\begin{itemize}
\item (Initial coloring, Version I) For all $k$-tuples $\overline{x}, \overline{y} \in G^k \cup H^k$, $\chi^{2,I}_0(\overline{x}) = \chi^{2,I}_0(\overline{y})$ iff $\overline{x}, \overline{y}$ are partially isomorphic.

\item (Initial coloring, Version II) For all $k$-tuples $\overline{x}, \overline{y} \in G^k \cup H^k$, $\chi^{2,II}_0(\overline{x}) = \chi^{2,II}_0(\overline{y})$ iff $\overline{x}, \overline{y}$ have the same marked isomorphism type.

\item (Color refinement) Given a coloring $\chi \colon G^k \cup H^k \to C$, the color refinement operator $R$ defines a new coloring $R(\chi)$ as follows. For each $k$-tuple $\overline{x} \in G^k$ (resp., $H^k$), we define an edge-colored graph $\Gamma_{\overline{x},\chi,i,j}$. If $i=j$, it is the graph on vertex set $V(\Gamma_{\overline{x},\chi,i,i}) = G$ (resp., $H$) with all self-loops and no other edges, where the color of each self-loop $(g,g)$ is $\chi(\overline{x}_{i \leftarrow g})$. If $i \neq j$, it is the complete directed graph with self-loops on vertex set $G$ (resp., $H$), where the color of each edge $(y,z)$ is $\chi(\overline{x}_{(i,j) \leftarrow (y,z)})$. For an edge-colored graph $\Gamma$, we use $[\Gamma]$ to denote its edge-colored isomorphism class. We then define
\[
R(\chi)(\overline{x}) = \left(\chi(\overline{x}); [\Gamma_{\overline{x},\chi,1,1}], [\Gamma_{\overline{x},\chi,1,2}], \dotsc, [\Gamma_{\overline{x},\chi,k,k-1}], [\Gamma_{\overline{x},\chi,k,k}]\right).
\]
That is, the new color consists of the old color, as well as the tuple of $\binom{k+1}{2}$ edge-colored isomorphism types of the graphs $\Gamma_{\overline{x}, \chi, i, j}$.
\end{itemize}
The refinement operator may be iterated: $R^t(\chi) := R(R^{t-1}(\chi))$, and we define the \emph{stable refinement} of $\chi$ as $R^t(\chi)$ where the partition induced by $R^t(\chi)$ on $G^k \cup H^k$ is the same as that induced by $R^{t+1}(\chi)$. We denote the stable refinement by $R^\infty(\chi)$.

Finally, for $J \in \{I, II\}$ and all $r \geq 0$, we define $\chi^{2,J}_{r+1} = R(\chi^{2,J}_r)$, and $\chi^{2,J}_{\infty} := R^\infty(\chi^{2,J}_0)$.
\end{defi}

\begin{rem}
Since it was one of our stumbling blocks in coming up with this generalized coloring, we clarify here how this indeed generalizes the usual 1-ary WL coloring procedure. In the 1-ary ``oblivious'' $k$-WL procedure (see \cite[\S 5]{grohe}, equivalent to ordinary WL), the color of a $k$-tuple $\overline{x}$ is refined using its old color, together with a $k$-tuple of multisets 
\[
(\{\!\{\chi(\overline{x}_{1 \leftarrow y}) : y \in G \}\!\}, \{\!\{\chi(\overline{x}_{2 \leftarrow y}) : y \in G \}\!\}, \dotsc, \{\!\{\chi(\overline{x}_{k \leftarrow y}) : y \in G \}\!\}).
\]
For each $i$, note that two multisets $\{\!\{ \chi(\overline{x}_{i \leftarrow y}) : y \in G \}\!\}$ and $\{\!\{ \chi(\overline{x}'_{i \leftarrow y}) : y \in G \}\!\}$ are equal iff the graphs $\Gamma_{\overline{x}, \chi, i, i}$ and $\Gamma_{\overline{x}', \chi, i, i}$ are color-isomorphic. That is, edge-colored graphs with only self-loops and no other edges are essentially the same, up to isomorphism, as multisets. Our procedure generalizes this by also considering graphs with other edges, which (as we'll see in the proof of equivalence below) are used to encode the choice of 2 simultaneous pebbles by Spoiler in each move of the game.
\end{rem}

\begin{thm} \label{thm:coloring}
Let $G, H$ be two groups of order $n$, with $\overline{x} = (x_1, \ldots, x_k) \in G^k$, $ \overline{y} = (y_1, \ldots, y_k) \in H^k$. Starting from the initial pebbling $x_i \mapsto y_i$ for all $i=1,\dotsc,k$, Spoiler has a winning strategy in the $k$-pebble, $r$-round, 2-ary Version $J$ pebble game (for $J \in \{I, II\}$) iff $\chi^{2,J}_r(\overline{x}) \neq \chi^{2,J}_r(\overline{y})$.
\end{thm}

\begin{proof}
By induction on $r$. The base case, $r=0$, is built into the definition: the initial colors of $\overline{x}, \overline{y}$ agree iff $x_i = x_j \Leftrightarrow y_i = y_j$ and in Version I, $x_i x_j = x_\ell \Leftrightarrow y_i y_j = y_\ell$, or in Version II the pebbled map is a marked isomorphism. But these are precisely the conditions for Spoiler to lose without making a move, so both directions are established for $r=0$.

Now suppose $r > 0$, and the equivalence is established for $r-1$. For ease of notation, throughout the proof we denote colorings $\chi^{2,J}$ by $\chi^J$ instead, and we denote the graphs $\Gamma_{\overline{x}, \chi^J_r, i, j}$ by $\Gamma_{\overline{x},r,i,j}$ instead ($q=2$ and the dependence on $J$ still being implied).

($\Leftarrow$) Suppose $\chi^J_r(\overline{x}) \neq \chi^J_r(\overline{y})$. If $\chi^J_{r-1}(\overline{x}) \neq \chi^J_{r-1}(\overline{y})$, then Spoiler has a winning strategy in $r-1$ rounds by the inductive hypothesis. Otherwise, we have $\chi^J_{r-1}(\overline{x}) = \chi^J_{r-1}(\overline{y})$. But since their colors differ at the $r$-th round, there must exist $i,j \in [k]$ such that $\Gamma_{\overline{x},r-1,i,j}$ is not color-isomorphic to $\Gamma_{\overline{y},r-1,i,j}$. Spoiler picks up the pebbles $i$ and $j$. Now, no matter what bijection $\varphi$ Duplicator chooses, $\varphi$ is not a color isomorphism on these graphs, so there exists an edge $(x_i', x_j') \in E(\Gamma_{\overline{x}, r-1, i, j})$ whose color differs from that of the edge $(\varphi(x_i'), \varphi(x_j')) \in E(\Gamma_{\overline{y}, r-1, i, j})$. (Note: there is no concern that the latter is not an edge of $\Gamma_{\overline{y},r-1,i,j}$, for $\varphi$ is automatically an uncolored graph isomorphism: when $i \neq j$ the graphs are complete, and when $i=j$ the graphs consist of $n$ isolated vertices with self-loops, so any bijection works.) Spoiler places the $i$-th pebble on $x_i' \mapsto \varphi(x_i') =: y_i'$ and the $j$-th pebble on $x_j' \mapsto \varphi(x_j') =: y_j'$. 

By the definition of the edge coloring, the fact that the edges $(x_i', x_j')$ and $(y_i', y_j')$ receive different colors in the $\Gamma_{\bullet, r-1, i, j}$ graphs means that $\chi^J_{r-1}(\overline{x}_{(i,j) \leftarrow (x_i', x_j')}) \neq \chi^J_{r-1}(\overline{y}_{(i,j) \leftarrow (y_i', y_j')})$. Now, by the inductive hypothesis applied to $\overline{x}_{(i,j) \leftarrow (x_i', x_j')}$ and $\overline{y}_{(i,j) \leftarrow (y_i', y_j')}$, Spoiler can win in at most $r-1$ additional rounds.

($\Rightarrow$) Suppose $\chi^J_r(\overline{x}) = \chi^J_r(\overline{y})$. We show that Duplicator has a strategy that does not lose through round $r$. On the first round, Spoiler picks up pebbles $i$ and $j$ (not necessarily distinct). Since $\chi^J_r(\overline{x}) = \chi^J_r(\overline{y})$, we have that $\Gamma_{\overline{x}, r-1, i, j}$ is color-isomorphic to $\Gamma_{\overline{y}, r-1, i, j}$, say by the isomorphism $\varphi \colon G = V(\Gamma_{\overline{x},r-1,i,j}) \to V(\Gamma_{\overline{y},r-1,i,j}) = H$. Duplicator uses the isomorphism $\varphi$ as their chosen bijection. Suppose Spoiler places the pebbles on $x_i' \mapsto \varphi(x_i') =: y_i'$ and $x_j' \mapsto \varphi(x_j') =: y_j'$. First, we claim that Duplicator has not yet lost. For the fact that $\varphi$ is a color isomorphism means that the colors of the edge $(x_i', x_j') \in E(\Gamma_{\overline{x},r-1,i,j})$ and $(y_i', y_j') \in E(\Gamma_{\overline{y},r-1,i,j})$ are the same. But these colors are precisely $\chi^J_{r-1}(\overline{x}_{(i,j) \leftarrow (x_i', x_j')})$ and $\chi^J_{r-1}(\overline{y}_{(i,j) \leftarrow (y_i', y_j')})$. Since $\chi^J_{r-1}$ refines $\chi^J_0$, we also have $\chi^J_0(\overline{x}_{(i,j) \leftarrow (x_i', x_j')}) = \chi^J_0(\overline{y}_{(i,j) \leftarrow (y_i', y_j')})$; since the 0-th coloring precisely matches the condition for Duplicator not to lose, Duplicator has not yet lost.

It remains to show that Duplicator can continue not to lose for an additional $r-1$ rounds. But this now follows from the inductive hypothesis applied to $\overline{x}_{(i,j) \leftarrow (x_i', x_j')}$ and $\overline{y}_{(i,j) \leftarrow (y_i', y_j')}$, for we have already established that they receive the same color under $\chi^J_{r-1}$.
\end{proof}

\begin{cor}
For two groups $G,H$ of the same order and any $k \geq 1$, the following are equivalent:
\begin{enumerate}
\item The 2-ary $k$-pebble game does not distinguish two groups $G,H$
\item The multisets of stable colors on $G^k$ and $H^k$ are the same, that is, $\{\!\{ \chi^{2,J}_\infty(\overline{x}) : \overline{x} \in G^k \}\!\} = \{\!\{ \chi^{2,J}_\infty(\overline{y}) : \overline{y} \in H^k\}\!\}$
\item $\chi^{2,J}_\infty(\underbrace{(1_G, 1_G, \dotsc, 1_G)}_{k}) = \chi^{2,J}_\infty(\underbrace{(1_H, \dotsc, 1_H)}_{k})$.
\end{enumerate}
\end{cor}

The analogous result holds in the $q=1$ case, going back to \cite{WLGroups}. 

\begin{proof}
($2 \Rightarrow 3$) Note that the identity tuples $(1, \dotsc, 1)$ are the unique tuples of their color, since they are the only tuples $(x_1, \dotsc, x_k)$ such that $x_i^2 = x_i$ for all $i$ (Version I), and the only tuples for which $\langle x_1, \dotsc, x_k \rangle = 1$ (Version II). Thus, if the multisets of stable colors are equal, the colors of $(1_G, \dotsc, 1_G)$ and $(1_H, \dotsc, 1_H)$ must be equal.

($3 \Rightarrow 2$) Suppose $\chi^{2,J}_{\infty}((1_G, \dotsc, 1_G)) = \chi^{2,J}_{\infty}((1_H, \dotsc, 1_H))$. Claim: for all $0 \leq \ell \leq k$, 
\begin{align*}
&\{\!\{ \chi^{2,J}_{\infty}(x_1, \dotsc, x_\ell, 1_G, \dotsc, 1_G) : x_1, \dotsc, x_\ell \in G \}\!\} \\
= &\{\!\{ \chi^{2,J}_{\infty}(x_1, \dotsc, x_\ell, 1_H, \dotsc, 1_H) : x_1, \dotsc, x_\ell \in H \}\!\}. \tag{$*$}\label{eq:claim}
\end{align*}
For notational simplicity, for $\overline{x} \in G^\ell$, define $\overline{x}^{+}$ as the $k$-tuple $(x_1, \dotsc, x_\ell, 1_G, \dotsc, 1_G)$, and similarly \emph{mutatis mutandis} for $\overline{y} \in H^\ell$. 

For $\ell=0$, (\ref{eq:claim}) is precisely our assumption (3).

Suppose (\ref{eq:claim}) is true for some $0 \leq \ell < k$. We show that it remains true for $\min\{\ell+2, k\}$. For notational simplicity, we assume $\ell+2 \leq k$; the proof in the case $k = \ell+1$ is similar.

By the inductive hypothesis, there is a bijection $\varphi_\ell \colon G^\ell \to H^\ell$ such that for all $\overline{x} \in G^\ell$,
\[
\chi^{2,J}_\infty(\overline{x}^+) = \chi^{2,J}_\infty(\varphi_\ell(\overline{x})^+)
\]
For each $\overline{x} \in G^\ell$, by the definition of the coloring, there is a color-isomorphism $\psi = \psi_{\overline{x}} \colon \Gamma_{\overline{x}^+, \infty, \ell+1, \ell+2} \to \Gamma_{\varphi(\overline{x})^+, \infty, \ell+1, \ell+2}$ (technically the subscripts here should be ``$\infty-1$,'' but because it is the stable coloring, the colored graphs are the same as what we have written). By the definition of the colors on the edges, this means that for all $x_{\ell+1}, x_{\ell+2} \in G$, we have that $\chi^J_\infty(x_1, \dotsc, x_\ell, x_{\ell+1}, x_{\ell+2}, 1_G, \dotsc, 1_G) = \chi^J_\infty(\varphi_\ell(\overline{x}), \psi(x_{\ell+1}), \psi(x_{\ell+2}), 1_H, \dotsc, 1_H)$ (where here we have slightly abused our parentheses, but in a way that should be clear). Thus, we may extend $\varphi_{\ell}$ to a color-preserving bijection $\varphi_{\ell+2}$ by defining
\[
\varphi_{\ell+2}(\overline{x}, x_{\ell+1}, x_{\ell+2}) = (\varphi_\ell(\overline{x}), \psi_{\overline{x}}(x_{\ell+1}), \psi_{\overline{x}}(x_{\ell+2})) \qquad \forall \overline{x} \in G^\ell, \, \forall x_{\ell+1}, x_{\ell+2} \in G.
\]
This establishes (\ref{eq:claim}) for $\ell+2$ (assuming $\ell+2 \leq k$). Consequently, this establishes the claim for all $\ell \leq k$, and thus that (3) implies (2).

($3 \Leftrightarrow 1$) The 2-ary $k$-pebble game starting from empty initial configurations is equivalent, move-for-move, with the 2-ary $k$-pebble game starting from the configuration \\ $(1_G,1_G,\dotsc,1_G) \mapsto (1_H,1_H, \dotsc, 1_H)$. By \Thm{thm:coloring}, the 2-ary $k$-pebble game thus does not distinguish $G$ from $H$ iff $\chi^{2,J}_{\infty}((1_G, \dotsc, 1_G)) = \chi^{2,J}_\infty((1_H, \dotsc, 1_H))$. 
\end{proof}

\begin{rem}
For arbitrary relational structures with relations of arity $q+1$, the $q$-ary pebble game may still be nontrivial, as pointed out in Hella \cite[p.~6, just before \S 4]{Hella1993}. Our coloring procedure generalizes in the following way to this more general setting, and the proof of the equivalence between the coloring procedure and Hella's pebble game is the same as the above, \emph{mutatis mutandis}. The main change is that for a $q$-ary pebble game, instead of just considering a graph on edges of size 1 (when $i=j$) or 2 (when $i \neq j$), we consider a $q'$-uniform directed hypergraph, where each hyperedge consists of a list of $q'$ vertices, for all $1 \leq q' \leq q$. This gives a coloring equivalent of the logical and game characterizations provided by Hella; this trifecta is partly why we feel it is justified to call this a ``higher-arity Weisfeiler--Leman'' coloring procedure. 

We note that there has been some work on equivalences with specific binary and higher-arity quantifiers---see for instance, the invertible map game of Dawar \& Holm \cite{DawarHolm} which generalizes rank logic, in which Spoiler can place multiple pebbles, but the bijections Duplicator selects must satisfy additional structure. Subsequently, Dawar \& Vagnozzi \cite{DawarVagnozzi} provided a generalization of Weisfeiler--Leman that further subsumes the invertible map game. We note that Dawar \& Vagnozzi's ``$WL_{k,r}$'', although it looks superficially like our $r$-ary $k$-WL, is in fact quite different: in particular, their refinement step ``flattens'' a multiset of multisets into its multiset union, which loses information compared to our 2-ary (resp., $r$-ary) game; indeed, they show that their WL$_{*,r}$ is equivalent to ordinary (1-ary) WL for any fixed $r$, whereas already 2-ary WL can solve GI. In general, the relationship between Hella's $2$-ary game and the works of Dawar \& Holm and Dawar \& Vagnozzi remains open. 
\end{rem}

\subsection{Equivalence between 2-ary $(k,r)$-WL Versions I and II} \label{sec:equiv12}
In this section we show that, up to additive constants in the number of pebbles and rounds, 2-ary WL Versions I and II are equivalent in their distinguishing power. For two different WL versions $W,W'$, we write $W \preceq W'$ to mean that if $W$ distinguishes two groups $G$ and $H$, then so does $W'$.

\begin{thm} \label{thm:rounds}
Let $k \geq 2, r \geq 1$. We have that:
\[
(k,r)\text{-WL}^2_{I} \preceq (k,r)\text{-WL}^2_{II} \preceq (k+2, r+1)\text{-WL}^2_{I}.
\]
\end{thm}

\begin{proof}
For the first statement, note that if $\overline{x} \mapsto \overline{y}$ is a marked isomorphism, then it is a partial isomorphism as well. After that, each color refinement step in the two versions are the same, so at whatever round Version I distinguishes $G$ from $H$, Version II will distinguish $G$ from $H$ at that round (or possibly sooner).

For the second statement, suppose that $(k,r)$-WL$_{II}^2$ distinguishes $G$ from $H$. Let $r_0 \leq r$ be the minimum number of rounds after which Spoiler has a winning strategy in this game. We show how Spoiler can win in the $(k+2, r_0+1)$-WL$_{I}^2$ game. For the first $r_0$ rounds, Spoiler makes exactly the same moves it would make in response to Duplicator's bijections in the Version II game. At the end of the $r_0$ round, Spoiler has pebbled a map $\overline{x} \mapsto \overline{y}$ of $k$-tuples that is not a marked isomorphism, but may still be a partial isomorphism. On the next round, Spoiler picks up the two pebbles $k+1, k+2$. Let $\varphi$ be the bijection Duplicator chooses. Since the pebbled map is not a marked isomorphism, there is a word $w$ such that $w(\varphi(x_1), \dotsc, \varphi(x_k)) \neq \varphi(w(\overline{x}))$. Let $w$ be such a word of minimal length $\ell$, and write $w(\overline{x}) = x_{i_1}^{\pm 1} w'(\overline{x})$, where $w'$ has length $\ell-1$. Spoiler places the two pebbles on $w'(\overline{x}) \mapsto \varphi(w'(\overline{x}))$ and $w(\overline{x}) \mapsto \varphi(w(\overline{x}))$. Now, since pebbles $i_1, k+1, k+2$ are on $x_{i_1}, w'(\overline{x}), w(\overline{x})$, resp., and $x_{i_1}^{\pm 1} w'(\overline{x}) = w(\overline{x})$ but $\varphi(x_i)^{\pm 1} \varphi(w'(\overline{x})) = w(\varphi(\overline{x})) \neq \varphi(w(\overline{x})$, Spoiler wins the Version I game.
\end{proof}

\section{Descriptive Complexity of Semisimple Groups} \label{SectionSemisimple}

In this section, we show that the $(O(1), O(1))$-WL$_{II}^{2}$ pebble game can identify groups with no Abelian normal subgroups, also known as semisimple groups. We begin with some preliminaries.

\subsection{Preliminaries} \label{SectionSemisimplePrelim}
Semisimple groups are motivated by the following characteristic\linebreak filtration:
\[
1 \leq \rad(G) \leq \Soc^{*}(G) \leq \text{PKer}(G) \leq G,
\]

\noindent which arises in the computational complexity community where it is known as the Babai--Beals filtration \cite{Babai1999GroupsSA}, as well as in the development of practical algorithms for computer algebra systems (cf., \cite{CH03}). We now explain the terms of this chain. Here, $\rad(G)$ is the \textit{solvable radical}, which is the unique maximal solvable normal subgroup of $G$. The socle of a group, denoted $\Soc(G)$, is the subgroup generated by all the minimal normal subgroups of $G$. $\Soc^{*}(G)$ is the preimage of the socle $\Soc(G/\rad(G))$ under the natural projection map $\pi : G \to G/\rad(G)$. To define $\text{PKer}$, we start by examining the action on $\Soc(G / \rad(G)) \cong \Soc^*(G) / \rad(G)$ that is induced by the action of $G$ on $\Soc^*(G)$ by conjugation. As $\Soc^{*}(G)/\rad(G) \cong \Soc(G/\rad(G))$ is the direct product of finite, non-Abelian simple groups $T_{1}, \ldots, T_{k}$, this action permutes the $k$ simple factors, yielding a homomorphism $\varphi : G \to S_{k}$. The kernel of this action is denoted $\text{PKer}(G)$.

When $\rad(G)$ is trivial, $G$ has no Abelian normal subgroups (and vice versa). We refer to such groups as \textit{semisimple} (following \cite{BCGQ, BCQ}) or trivial-Fitting (following \cite{CH03}). As a semisimple group $G$ has no Abelian normal subgroups, we have that $\Soc(G)$ is the direct product of non-Abelian simple groups.  As the conjugation action of $G$ on $\Soc(G)$ permutes the direct factors of $\Soc(G)$, there exists a faithful permutation representation $\alpha : G \to G^{*} \leq \Aut(\Soc(G))$. $G$ is determined by $\Soc(G)$ and the action $\alpha$. Let $H$ be a semisimple group with the associated action $\beta : H \to \text{Aut}(\Soc(H))$. We have that $G \cong H$ precisely if $\Soc(G) \cong \Soc(H)$ via an isomorphism that makes $\alpha$ equivalent to $\beta$ in the sense introduced next. 

We now introduce the notion of permutational isomorphism, which is our notion of equivalence for $\alpha$ and $\beta$. Let $A$ and $B$ be finite sets, and let $\pi : A \to B$ be a bijection. For $\sigma \in \text{Sym}(A)$, let $\sigma^{\pi} \in \text{Sym}(B)$ be defined by $\sigma^{\pi} := \pi^{-1}\sigma \pi$. For a set $\Sigma \subseteq \text{Sym}(A)$, denote $\Sigma^{\pi} := \{ \sigma^{\pi} : \sigma \in \Sigma\}$. Let $K \leq \text{Sym}(A)$ and $L \leq \text{Sym}(B)$ be permutation groups. A bijection $\pi : A \to B$ is a \textit{permutational isomorphism} $K \to L$ if $K^{\pi} = L$.

The following lemma, applied with $R = \Soc(G)$ and $S = \Soc(H)$, gives a precise characterization of semisimple groups in terms of the associated actions.      
 
\begin{lemC}[{\cite[Lemma 3.1]{BCGQ}, cf. \cite[\S 3]{CH03}}] \label{CharacterizeSemisimple}
Let $G$ and $H$ be groups, with $R \triangleleft G$ and $S \triangleleft H$ groups with trivial centralizers. Let $\alpha : G \to G^{*} \leq \Aut(R)$ and $\beta : H \to H^{*} \leq \Aut(S)$ be faithful permutation representations of $G$ and $H$ via the conjugation action on $R$ and $S$, respectively. Let $f : R \to S$ be an isomorphism. Then $f$ extends to an isomorphism $\hat{f} : G \to H$ if and only if $f$ is a permutational isomorphism between $G^{*}$ and $H^{*}$; and if so, $\hat{f} = \alpha f^{*} \beta^{-1}$, where $f^{*} :  G^{*} \to H^{*}$ is the isomorphism induced by $f$.
\end{lemC}

\begin{rem}
We note that \Lem{CharacterizeSemisimple} depends only on the assumption that the centralizers $C_{G}(R)$ and $C_{H}(S)$ are trivial. In particular, \Lem{CharacterizeSemisimple} does not depend on the Classification of Finite Simple Groups (CFSG); the only dependence our results have on CFSG is through its consequence that all finite simple groups are 2-generated.
\end{rem}

We also need the following standard group-theoretic lemmas. The first provides a key condition for identifying whether a non-Abelian simple group belongs in the socle. Namely, if $S_{1} \cong S_{2}$ are non-Abelian simple groups where $S_{1}$ is in the socle and $S_{2}$ is not in the socle, then the normal closures of $S_{1}$ and $S_{2}$ are non-isomorphic. In particular, the normal closure of $S_{1}$ is a direct product of non-Abelian simple groups, while the normal closure of $S_{2}$ is not a direct product of non-Abelian simple groups. We will apply this condition later when $S_{1}$ is a simple direct factor of $\Soc(G)$; in which case, the normal closure of $S_{1}$ is of the form $S_{1}^{k}$. We include the proofs of these two lemmas for completeness.

\begin{lem} \label{LemmaSocle}
Let $G$ be a finite semisimple group. A subgroup $S \leq G$ is contained in $\Soc(G)$ if and only if the normal closure of $S$ is a direct product of non-Abelian simple groups.
\end{lem}

\begin{proof}
Let $N$ be the normal closure of $S$. Since the socle is normal in $G$ and $N$ is the smallest normal subgroup containing $S$, we have that $S$ is contained in $\Soc(G)$ if and only if $N$ is. 

Suppose first that $S$ is contained in the socle. Since $\Soc(G)$ is normal and contains $S$, by the definition of $N$ we have that $N \leq \Soc(G)$. As $N$ is a normal subgroup of $G$, contained in $\Soc(G)$, it is a direct product of minimal normal subgroups of $G$, each of which is a direct product of non-Abelian simple groups.

Conversely, suppose $N$ is a direct product of non-Abelian simple groups. We proceed by induction on the size of $N$. If $N$ is minimal normal in $G$, then $N$ is contained in the socle by definition. If $N$ is not minimal normal, then it contains a proper subgroup $M \lneq N$ such that $M$ is normal in $G$, hence also $M \unlhd N$. However, as $N$ is a direct product of non-Abelian simple groups $T_1, \dotsc, T_k$, the only subgroups of $N$ that are normal in $N$ are direct products of subsets of $\{T_1, \dotsc, T_k\}$, and all such normal subgroups have direct complements. Thus we may write $N = L \times M$ where both $L,M$ are nontrivial, hence strictly smaller than $N$, and both $L$ and $M$ are direct product of non-Abelian simple groups. 

We now argue that $L$ must also be normal in $G$. Since conjugating $N$ by $g \in G$ is an automorphism of $N$, we have that $N = g L g^{-1} \times g M g^{-1}$. Since $M$ is normal in $G$, the second factor here is just $M$, so we have $N = gLg^{-1} \times M$. But since the direct complement of $M$ in $N$ is unique (since $N$ is a direct product of \emph{non-Abelian} simple groups), we must have $g L g^{-1} = L$. Thus $L$ is normal in $G$. 

By induction, both $L$ and $M$ are contained in $\Soc(G)$, and thus so is $N$. We conclude since $S \leq N$.
\end{proof}

\begin{lem} \label{LemmaDirectProdSimple}
Let $S_1, \dotsc, S_k \leq G$ be non-Abelian simple subgroups such that for all distinct $i,j \in [k]$ we have $[S_i, S_j] = 1$. Then $\langle S_1, \dotsc, S_k \rangle = S_1 S_2 \dotsb S_k = S_1 \times \dotsb \times S_k$.
\end{lem}

\begin{proof}
By induction on $k$. The base case $k=1$ is vacuously true. Suppose $k \geq 2$ and that the result holds for $k-1$. Then $T := S_1 S_2 \dotsb S_{k-1} = S_1 \times \dotsb \times S_{k-1}$. Now, since $S_k$ commutes with each $S_i$, and they generate $T$, we have that $[S_k, T] = 1$. Hence $T$ is contained in the normalizer (or even the centralizer) of $S_k$, so $TS_k = S_kT = \langle T, S_k \rangle$, and $S_k$ and $T$ are normal subgroups of $TS_k$. As $TS_k = \langle T,S_k \rangle$ and $T,S_k$ are both normal subgroups of $TS_k$ with $[T, S_k] = 1$, we have that $TS_k$ is a central product of $T$ and $S_k$. As $Z(T) = Z(S_k)=1$, it is their direct product.
\end{proof}

\subsection{Main Results}
We show that the second Ehrenfeucht--Fra\"iss\'e game in Hella's hierarchy can identify both $\Soc(G)$ and the conjugation action when $G$ is semisimple. We first show that this pebble game can identify whether a group is semisimple. Namely, if $G$ is semisimple and $H$ is not semisimple, then Spoiler can distinguish $G$ from $H$. 

\begin{prop} \label{IdentifySemisimple}
Let $G$ be a semisimple group of order $n$, and let $H$ be an arbitrary group of order $n$. If $H$ is not semisimple, then Spoiler can win in the $(3,2)$-WL$_{II}^{2}$ game.
\end{prop}

\begin{proof}
Recall that a group is semisimple if and only if it contains no Abelian normal subgroups. As $H$ is not semisimple, $\Soc(H) = A \times T$, where $A$ is a nontrivial subgroup that is the direct product of elementary Abelian groups and $T$ is a direct product of non-Abelian simple groups. We show that Spoiler can win using at most $3$ pebbles and $2$ rounds. Let $f : G \to H$ be the bijection that Duplicator selects. Let $a \in A$. So $\text{ncl}_{H}(a) \leq A$. Let $b := f^{-1}(a) \in G$, and let $B := \text{ncl}_{G}(b)$. As $G$ is semisimple, we have that $B$ is not Abelian. Spoiler first pebbles $b \mapsto f(b) = a$.

So there exist $g_{1}, g_{2} \in G$ such that $g_{1}bg_{1}^{-1}$ and $g_{2}bg_{2}^{-1}$ do not commute (for $B$ is generated by $\{g b g^{-1} : g \in G\}$, and if they all commuted then $B$ would be Abelian). Let $f^{\prime} : G \to H$ be the bijection that Duplicator selects at the next round. If $f'(g_i^{-1}) \neq f'(g_i)^{-1}$ for either $i=1,2$, Spoiler uses the two pebbles in hand to win that round, by Observation~\ref{obs:inverse}, for a total of 3 pebbles and 2 rounds.

Otherwise, Spoiler pebbles $(g_{1}, g_{2}) \mapsto (f^{\prime}(g_{1}), f^{\prime}(g_{2}))$. As $\text{ncl}(f(b)) \leq A$ is Abelian, $f^{\prime}(g_{1})f(b)f^{\prime}(g_{1})^{-1}$ and $f^{\prime}(g_{2})f(b)f^{\prime}(g_{2})^{-1}$ commute. Spoiler now wins, again with a total of 3 pebbles and 2 rounds.
\end{proof}
 
We now apply Lemma \ref{LemmaSocle} to show that Duplicator must map the direct factors of $\Soc(G)$ to isomorphic direct factors of $\Soc(H)$. 

\begin{lem} \label{LemmaProdSimple}
Let $G,H$ be finite groups of order $n$. Let $\Fac(\Soc(G))$ denote the set of simple direct factors of $\Soc(G)$. Let $S \in \Fac(\Soc(G))$ be a non-Abelian simple group, with $S = \langle x, y \rangle$. If Duplicator selects a bijection $f \colon G \to H$ such that:
\begin{enumerate}[label=(\alph*)]
\item $S \not \cong \langle f(x), f(y) \rangle$, then Spoiler can win in the $(2,1)$-WL$_{II}^{2}$ game; or 
\item $f(S) \neq \langle f(x), f(y) \rangle$, then Spoiler can win in the $(3,2)$-WL$_{II}^{2}$ pebble game.
\end{enumerate}
\end{lem}

\noindent \\ Note that Lemma~\ref{LemmaProdSimple}(b) stipulates that, in order not to lose, Duplicator must map the elements in $S$ precisely to the elements in $\langle f(x), f(y) \rangle$. Furthermore, $S$ and $\langle f(x), f(y) \rangle$ must be isomorphic. However, in order to avoid losing, this lemma does not yet establish that it is necessary for Duplicator to choose a bijection $f$ such that the restriction $f|_{S} : S \to f(S)$ is an isomorphism.

\pagebreak
\begin{proof}
\noindent
\begin{enumerate}[label=(\alph*)]
\item If $S \not \cong \langle f(x), f(y) \rangle$, then Spoiler pebbles $(x, y) \mapsto (f(x), f(y))$ and immediately wins.

\item Suppose that $S \cong \langle f(x), f(y) \rangle$. By part (a), we may assume that the map $(x, y) \mapsto (f(x), f(y))$ extends to an isomorphism; otherwise Spoiler immediately wins. So suppose that $f(S) \neq \langle f(x), f(y) \rangle$. Note that as $S \cong \langle f(x), f(y) \rangle$, we have that $|S| = |\langle f(x), f(y) \rangle|$. So $f(S) \subseteq \langle f(x), f(y) \rangle$ if and only if $f(S) = \langle f(x), f(y) \rangle$. 

Now by the assumption that $f(S) \neq \langle f(x), f(y) \rangle$, there exists an element $b \in S$ such that $f(b) \not \in \langle f(x), f(y) \rangle$. Let $a \in S$ such that $a \neq 1$ and $f(a) \in \langle f(x), f(y) \rangle$. Spoiler pebbles $(a, b) \mapsto (f(a), f(b))$. Let $f' : G \to H$ be the bijection that Duplicator chooses on the next round. Let $g, h \in G$ such that $f'(g) = f(x)$ and $f'(h) = f(y)$. Spoiler now pebbles $(g, h) \mapsto (f(x), f(y))$. As $S \cong \langle f(x), f(y) \rangle$ by assumption, if $T = \langle g, h \rangle$ is not isomorphic to $S$, then Spoiler wins immediately. Furthermore, as $f(a) \in \langle f(x), f(y) \rangle$, we may assume that $a \in \langle g, h \rangle$; otherwise, Spoiler immediately wins. 

We now claim that $T \leq \Soc(G)$. As $\Soc(G)$ is normal in $G$, we have that $T \cap \Soc(G)$ is normal in $T$. As $T$ is simple and intersects with $\Soc(G)$ non-trivially (namely, $a \in T \cap \Soc(G)$), it follows that $T \leq \Soc(G)$. Now as $S \trianglelefteq \Soc(G)$ and $T \leq \Soc(G)$, we have by similar argument (using the fact that $a \in S$) that $T = S$. So $S = \langle g, h \rangle$. Now $a, b \in S = \langle g, h \rangle$; however, $f(b) \not \in \langle f(x), f(y) \rangle$. So Spoiler wins. \qedhere
\end{enumerate}
\end{proof}

\begin{prop} \label{PropSocleSemisimple}
Let $G, H$ be semisimple groups of order $n$. Let $f : G \to H$ be the bijection Duplicator selects. If there exists $S \in \Fac(\Soc(G))$ such that $f(S) \notin \Fac(\Soc(H))$ or $f(S) \not\cong S$, then Spoiler can win in the $(4,3)$-WL$_{II}^{2}$ pebble game.
\end{prop}

\begin{proof}
As $G$ is semisimple, we have that $S = \langle x, y \rangle$ is non-Abelian. Let $f : G \to H$ be the bijection Duplicator selects. We may assume that $f(S) \cong S$ (though $f|_{S}$ need not be an isomorphism); otherwise, Spoiler can win with $2$ pebbles and $1$ round by pebbling the generators $(x, y)$ for $S$. Furthermore, we may assume that $f(S) = \langle f(x), f(y) \rangle$ setwise; otherwise, Spoiler can win with $3$ pebbles and $2$ rounds by \Lem{LemmaProdSimple}. 

Suppose that $f(S)$ is not a direct factor of $\Soc(H)$. Spoiler pebbles $(x, y) \mapsto (f(x), f(y))$. We now have the following cases.

\begin{itemize}
\item \textbf{Case 1:} Suppose that $f(S)$ is not contained in $\Soc(H)$. Let $f' : G \to H$ be the bijection that Duplicator selects at the next round. As $(x,y) \mapsto (f(x), f(y))$ have been pebbled, if $f'|_S$ is not an isomorphism, then Spoiler wins with 1 additional pebble and 1 additional round, for a total of 3 pebbles and 2 rounds: for, since $\{x,y\}$ generates $S$, if $f'|_S$ is not an isomorphism then there is some $z \in S$ such that $(x,y,z) \mapsto (f(x), f(y), f'(z))$ does not extend to an isomorphism, and Spoiler can pebble such a $z$.

Next, as $f(S) = \langle f(x), f(y) \rangle$, we must have that at least one of $f(x), f(y)$ is not in $\Soc(H)$, and therefore $f'(S) \supseteq \{ f(x), f(y) \}$ is also not contained in $\Soc(H)$. 
As $S \triangleleft \text{Soc}(G)$, the normal closure $\text{ncl}(S)$ is minimal normal in $G$ \cite[Exercise 2.A.7]{Isaacs2008FiniteGT}. But since $f'(S)$ is not even contained in $\text{Soc}(H)$, we have by Lemma \ref{LemmaSocle} that $\text{ncl}(f'(S))$ is not a direct product of non-Abelian simple groups, so $\text{ncl}(S) \not\cong \text{ncl}(f'(S))$. We note that $\text{ncl}(S) = \langle \{ gSg^{-1} : g \in G \} \rangle$.

As $\text{ncl}(f'(S))$ is not isomorphic to a direct power of $S$, there is some conjugate $T = g S g^{-1} \neq S$ such that $f'(T)$ does not commute with $f'(S)$, by Lemma~\ref{LemmaDirectProdSimple}. Yet since $S \unlhd \Soc(G)$, $T$ and $S$ do commute. Spoiler pebbles $g \mapsto f'(g)$. Spoiler has now pebbled $x,y,g$ which generate a subgroup containing $\langle S, T \rangle = S \times T \leq G$ and $S \times T \cong S \times S$.  However, as the image is not isomorphic to $S \times S$, the map $(x,y,g) \mapsto (f(x), f(y), f'(g))$ does not extend to an isomorphism of $\langle x,y,g \rangle$ with $\langle f(x), f(y), f'(g) \rangle$, as any such isomorphism would have to map $S \times T \leq \langle x, y, g \rangle$ to a subgroup of $H$ isomorphic to $S \times S$. So Spoiler now wins, with a total of $3$ pebbles and $2$ rounds.

\item \textbf{Case 2:} Suppose now that $f(S) \leq \Soc(H)$, but that $f(S)$ is not normal in $\Soc(H)$. As $f(S)$ is not normal in $\Soc(H)$, there exists $T = \langle a, b \rangle \in \Fac(\Soc(H))$ such that $T$ does not normalize $f(S)$. 
Let $f' : G \to H$ be the bijection that Duplicator selects at the next round. Again, we may assume $f'|_{S}$ is an isomorphism. By \Lem{LemmaProdSimple}, we may assume without loss of generality that for each $T \in \Fac(\Soc(G))$, $f'(T) \cong T$ (though $f'|_{T}$ need not be an isomorphism) and, by the previous case, that $f'(T) \leq \Soc(H)$. Otherwise, Spoiler wins with $2$ additional pebbles and $2$ additional rounds. Note that while Case 1 prescribes $3$ additional pebbles, we may at the next round reuse pebbles on $x,y$, and only one such pebble is needed at the next round. This brings our total down to $2$ additional pebbles.
Let $g = (f')^{-1}(a)$ and $h = (f')^{-1}(b)$. If either $g$ or $h$ is not in $\Soc(G)$, then by \Lem{LemmaProdSimple}(b), Spoiler can use the two pebbles in hand this round, plus one additional pebble and one additional round to win, for a total of 4 pebbles and 3 rounds. Otherwise, as $S$ is normal in $\Soc(G)$, we note that $gSg^{-1} = S$ and $hSh^{-1} = S$.  Spoiler pebbles $(g, h) \mapsto (a, b)$. The map $(x, y, g, h) \mapsto (f(x), f(y), a, b)$ does not extend to an isomorphism of $\langle x, y, g, h \rangle$ and $\langle f(x), f(y), a, b \rangle$, since $g,h$ normalize $S = \langle x, y \rangle$ but $a,b$ cannot both normalize $f(S) = \langle f(x), f(y) \rangle$ (since $T$ does not). Spoiler now wins. In this case, we have used at most $4$ pebbles and $3$ rounds.
\end{itemize} 

The result follows.
\end{proof}

\begin{lem} \label{LemmaSimpleOverlap}
Let $G,H$ be semisimple groups of order $n$, let $S$ be a non-Abelian simple group in $\Fac(\Soc(G))$. Let $f,f'\colon G \to H$ be two bijections selected by Duplicator at two different rounds. If $f(S) \cap f'(S) \neq 1$, then $f(S) = f'(S)$, or Spoiler can win in the $(4,3)$-WL$_{II}^{2}$ pebble game.
\end{lem}

\begin{proof}
By \Prop{PropSocleSemisimple}, both $f(S)$ and $f'(S)$ must be simple normal subgroups of $\Soc(H)$ (or Spoiler wins with $4$ pebbles and $3$ rounds). Since they intersect nontrivially, but distinct simple normal subgroups of $\Soc(H)$ intersect trivially, the two must be equal. 
\end{proof}

We next introduce the notion of weight.

\begin{defi}
Let $\Soc(G) = S_1 \times \dotsb \times S_k$ where each $S_i$ is a simple normal subgroup of $\Soc(G)$. For any $s \in \Soc(G)$, write $s = s_1 s_2 \dotsb s_k$ where each $s_i \in S_i$, and define the \emph{weight} of $s$, denoted $\wt(s)$, as the number of $i$'s such that $s_i \neq 1$. For $s \notin \Soc(G)$ we define $\wt(s) = \infty$.
\end{defi}

Note that the definition of weight is well-defined since the $S_i$ are the unique subsets of $\Soc(G)$ that are simple normal subgroup of $\Soc(G)$, so the decomposition $s = s_1 s_2 \dotsc s_k$ is unique up to the order of the factors.\footnote{This is essentially a particular instance of the ``rank lemma'' from \cite{GrochowLevetWL}, which intuitively states that WL detects in $O(\log n)$ rounds the set of elements for a given subgroup provided that it also identifies the generators. As we are now in the setting of $2$-ary WL we give the full proof, which also has tighter bounds on the number of rounds.}

\begin{lem}[Weight Lemma] \label{LemmaSemisimpleWeight}
Let $G, H$ be semisimple groups of order $n$. If Duplicator selects a bijection that does not preserve the weight of every element of $G$, then Spoiler can win in the $(4,4)$-WL$_{II}^{2}$ game. In particular, Spoiler can win the $(4,4)$-WL$_{II}^{2}$ game if Duplicator selects a bijection that does not map $\Soc(G)$ bijectively to $\Soc(H)$.
\end{lem}

\begin{proof}
The identity is the unique element of weight 0, which must be sent to the identity because it is the unique element $e$ satisfying $e^2 = e$. Otherwise Spoiler can win by pebbling the identity with 1 pebble in 1 round. The case of weight 1 is precisely Proposition~\ref{PropSocleSemisimple}, which needs only 4 pebbles and 3 rounds.

Suppose for the sake of contradiction that $f$ does not preserve weight. Among the elements $s \in G$ such that $\wt(f(s)) \neq \wt(s)$, choose one of minimal weight $w$. We must have $w \geq 2$, since we have already handled the cases of weight 0 and 1. Furthermore, we may assume $w$ is finite; if $|\Soc(G)| \neq |\Soc(H)|$, then we may choose $G$ to be the one with larger socle, whereby some finite-weight element must get sent to an infinite-weight element, and if the socles have the same size, then since $f$ is a bijection, if it doesn't preserve the weight of some infinite-weight element, it must also map some finite-weight element to an infinite-weight element.

Since $f$ is a bijection from the elements of weight $< w$ in $\Soc(G)$ to the elements of weight $< w$ in $\Soc(H)$ (by the choice of $s$), and $\wt(f(s)) \neq w$, we must have $\wt(f(s)) > w$ since $f$ is a bijection. 

Without loss of generality, write $\Soc(G) = S_1 \times \dotsb \times S_k$ where the $S_i$ are the simple normal subgroups of $\Soc(G)$ and $s \in S_1 \times S_2 \times \dotsb \times S_w$, where we can write $s = s_1 s_2 \dotsb s_w$ with all $s_i \neq 1$. Since $s$ is a smallest-weight element whose weight is not preserved, $f(s_2 s_3 \dotsb s_w)$ must have weight precisely $w-1$. Now $f(s_1)$ has weight 1, so $f(s_1) f(s_2 s_3 \dotsb s_w)$ has weight either $w-2$, $w-1$ or $w$. But since $\wt(f(s)) > w$, we have $f(s_1 s_2 \dotsb s_w) = f(s) \neq f(s_1) f(s_2 s_3 \dotsb s_w)$, since the two elements have different weights. Now Spoiler pebbles the pair $(s, s_2 s_3 \dotsb s_w) \mapsto (f(s), f(s_2 \dotsb s_w))$. 

At the next round, Duplicator selects another bijection $f'$. We still have $\wt(f'(s)) = \wt(f(s)) > w$, $\wt(f'(s_2 \dotsb s_w)) = \wt(f(s_2 \dotsb s_w)) = w-1$, and (by Proposition~\ref{PropSocleSemisimple}) $\wt(f'(s_1)) = 1$. Note that although Proposition~\ref{PropSocleSemisimple} uses 4 pebbles and 3 rounds, after the first round of that protocol, Spoiler may reuse the pebbles that were previously on $s$ and $s_2 s_3 \dotsb s_w$. Since the first round of that protocol was the second round in this proof, the total here would be 4 pebbles and 4 rounds. Thus, again, we have $\wt(f'(s_1) f'(s_2 s_3 \dotsb s_w)) \in \{w,w-1, w-2\}$, so we must have $f'(s_1 s_2 \dotsb s_w) = f'(s) \neq f'(s_1) f'(s_2 \dotsb s_w)$. Spoiler now pebbles $s_1$ and wins with $4$ pebbles and $2$ rounds. In either case, Spoiler used at most 4 pebbles and at most 4 rounds.
\end{proof}

\begin{lem} \label{SocleDirectProductStronger}
Let $G$ and $H$ be semisimple groups with isomorphic socles. Let $S_{1}, S_{2} \in \Fac(\Soc(G))$ be distinct. Let $f : G \to H$ be the bijection that Duplicator selects. If there exist $x_i \in S_i$ such that $f(x_1 x_2) \neq f(x_1) f(x_2)$, then Spoiler can win in the $(4,5)$-WL$_{II}^{2}$ pebble game. 
\end{lem}

\begin{proof}
By \Lem{LemmaSemisimpleWeight}, we may assume that $\wt(s) = \wt(f(s))$ for all $s \in \Soc(G)$; otherwise, Spoiler wins with at most $4$ pebbles and $4$ rounds. As $f(x_{1}x_{2})$ has weight $2$, $f(x_{1}x_{2})$ belongs to the direct product of two simple factors in $\Fac(\Soc(H))$, so it can be written $f(x_1 x_2) = y_1 y_2$ with each $y_i$ in distinct simple factors in $\Fac(\Soc(H))$. Without loss of generality suppose that $y_1 \neq f(x_1)$. Spoiler pebbles $(x_1, x_1 x_2) \mapsto (f(x_1), f(x_1 x_2))$. Now $\wt(x_1^{-1} \cdot x_{1}x_{2})  = 1$, while $\wt(f(x_1)^{-1} \cdot f(x_{1}x_{2})) \geq 2$. (Note that we cannot quite yet directly apply \Lem{LemmaSemisimpleWeight}, because we have not yet identified a single element $x$ such that $\wt(x) \neq \wt(f(x))$.)

On the next round, Duplicator selects another bijection $f'$. Spoiler now pebbles $x_2 \mapsto f'(x_2)$. Suppose that $f'$ preserves weight. Otherwise,  by \Lem{LemmaSemisimpleWeight}, Spoiler wins with $4$ pebbles, reusing existing pebbles on the board, and $4$ additional rounds for a total of $5$ rounds. Because $\wt(x_1^{-1} \cdot x_1 x_2) = 1$ but $\wt(f(x_1)^{-1} f(x_1 x_2)) \geq 2$, and $f'$ preserves weight, we have $f'(x_2) \neq f'(x_1)^{-1} f'(x_1 x_2)$. Thus, the pebbled map $(x_1, x_2, x_1 x_2) \mapsto (f'(x_1), f(x_2), f(x_1 x_2))$ does not extend to an isomorphism, and so Spoiler wins with $3$ pebbles and $2$ rounds. 

In total, Spoiler used at most $4$ pebbles and at most $5$ rounds.
\end{proof}

Recall that if $G$ is semisimple, then $G \cong G^{*} \leq \Aut(\Soc(G))$ (see Section~\ref{SectionSemisimplePrelim}). Now each minimal normal subgroup $N \trianglelefteq G$ is of the form $N = S^{k}$, where $S$ is a non-Abelian simple group. So $\Aut(N) = \Aut(S) \wr \text{Sym}(k)$. In particular, 
\[
G \leq \prod_{ \substack{ N \trianglelefteq G \\ N \text{ is minimal normal}} } \Aut(N).
\]

So if $g \in G$, then the conjugation action of $g$ on $\Soc(G)$ acts by (i) automorphism on each simple direct factor of $\Soc(G)$, and (ii) by permuting the direct factors of $\Soc(G)$. Provided generators of the direct factors of the socle are pebbled, Spoiler can detect inconsistencies of the automorphism action. However, doing so directly would be too expensive as there could be $\Theta(\log|G|)$ generators, so we employ a more subtle approach with a similar outcome. By \Lem{LemmaSemisimpleWeight}, Duplicator must select bijections $f : G \to H$ that preserve weight. 
We use \Lem{LemmaSemisimpleWeight} in tandem with the fact that the direct factors of the socle commute to effectively pebble the set of all the generators at once. Namely, suppose that $\Fac(\Soc(G)) = \{ S_{1}, \ldots, S_{k}\}$, where $S_{i} = \langle x_{i}, y_{i} \rangle$. Let $x := x_{1} \cdots x_{k}$ and $y := y_{1} \cdots y_{k}$. We will show that it suffices for Spoiler to pebble $(x, y)$ rather than individually pebbling generators for each $S_{i}$ (this will still allow the factors to be permuted, but that is all).

\begin{lem} \label{SemisimpleFactors}
Let $G$ and $H$ be semisimple groups with isomorphic socles, with $\Fac(\Soc(G)) = \{S_1, \dotsc, S_m\}$, with $S_i = \langle x_i, y_i \rangle$. Let $f : G \to H$ be the bijection that Duplicator selects, and suppose that (i) for all $i$, $f(S_i) \cong S_i$ (though $f|_{S_i}$ need not be an isomorphism) and $f(S_i) \in \Fac(\Soc(H))$, (ii) for every $s \in G$, $\wt(s) = \wt(f(s))$, and
(iii) for all $i$, $f(S_i) = \langle f(x_i), f(y_i) \rangle$. 

Now suppose that Spoiler pebbles $(x_{1} \cdots x_{m}, y_{1} \cdots y_{m}) \mapsto (f(x_{1} \cdots x_{m}), f(y_{1} \cdots y_{m}))$. As $f$ preserves weight, we may write $f(x_{1} \cdots x_{m}) = h_{1} \cdots h_{m}$ and $f(y_{1} \cdots y_{m}) = z_{1} \cdots z_{m}$ with $h_i, z_i \in f(S_i)$ for all $i$.

Let $f' : G \to H$ be the bijection that Duplicator selects at any subsequent round in which the pebbles used above have not moved. 
If any of the following hold, then Spoiler can win in the WL$_{II}^{2}$ pebble game with  $4$ additional pebbles and $5$ additional rounds (beyond the number of current pebbles and rounds, when this lemma is applied):
\begin{enumerate}[label=(\alph*)]
\item $f'$ does not satisfy conditions (i)--(iii),
\item there exists an $i \in [m]$ such that $f'(x_i) \notin \{h_1, \dotsc, h_m\}$ or $f'(y_i) \notin \{z_1, \dotsc, z_m\}$
\item there exists an $i \in [m]$ such that $f'|_{S_{i}}$ is not an isomorphism
\item there exists $g \in G$ and $i \in [m]$ such that $gS_{i}g^{-1} = S_{i}$ and for some $x \in S_{i}$, $f'(gxg^{-1}) \neq f'(g)f'(x)f'(g)^{-1}$
\end{enumerate}

Furthermore, cases (a) and (b) instead only require $4$ additional pebbles and $4$ additional rounds (rather than $5$ additional rounds).
\end{lem}

\begin{proof}
We prove each part in turn:
\begin{enumerate}[label=(\alph*)]
\item If $f'$ does not satisfy (i)--(iii) then Spoiler can win with $4$ additional pebbles and $4$ additional rounds by, respectively, \Prop{PropSocleSemisimple}, \Lem{LemmaSemisimpleWeight}, and \Lem{LemmaProdSimple}.

\item Suppose there exists an $i \in [m]$ such that for all $j \in [m]$, $f'(x_{i}) \neq h_{j}$. Spoiler pebbles $(x_i, x_1 \dotsb x_m x_i^{-1}) \mapsto (f'(x_i), f'(x_1 \dotsb x_m x_i^{-1}))$. 
Now $\wt(x_{1} \cdots x_{m} \cdot x_{i}^{-1}) = m-1$, while $\wt(f(x_{1} \cdots x_{m}) \cdot f'(x_{i})^{-1}) = m$, since $f'(x_i) \notin \{h_1, \dotsc, h_m\}$. Since $f'$ preserves weight (by part (a)), $f'(x_1 \dotsb x_m) f'(x_i)^{-1} \neq f'(x_1 \dotsb x_m x_i^{-1})$. As all three of these elements have been pebbled, Spoiler wins. Similarly if instead some $y_i$ has $f'(y_i) \notin \{z_1, \dotsc, z_m\}$. In total, Spoiler used at most $4$ additional pebbles and $4$ additional rounds (by part (a)).

\item First, by (b), we may assume that $f'(x_i) = h_j$ and $f'(y_i) = z_j$ for some $j$. Otherwise, Spoiler wins with $4$ additional pebbles and $4$ additional rounds. Now suppose that for some $i \in [k]$, $f'|_{S_{i}}$ is not an isomorphism. Then there is some word $w(x,y)$ such that $f'(w(x_i, y_i)) \neq w(f'(x_i), f'(y_i))$. We may further assume $w(x_i, y_i) \neq 1$, since if $f'$ preserves all such words, then $f'$ must preserve \emph{all} words and thereby be an isomorphism. 

Spoiler pebbles $g = w(x_i, y_i)$. Let $f'' : G \to H$ be the bijection that Duplicator selects at the next round. Since $f'(S_i)$ and $f''(S_i)$ intersect at $f''(g) \neq 1$, by \Lem{LemmaSimpleOverlap}, we have $f'(S_i) = f''(S_i)$; otherwise, Spoiler wins with $2$ additional pebbles and $3$ additional rounds. Note that while \Lem{LemmaSimpleOverlap} prescribes $4$ additional pebbles, we may reuse the two pebbles on $x_{1} \cdots x_{m}$ and $y_{1} \cdots y_{m}$.
 
Recall from the first paragraph in the proof for (c), we have that $f'(x_i) = h_j$ and $f'(y_i) = z_j$ for some $j$. Since $f''(x_i) \in f''(S_i) = f'(S_i)$, and by (b) we must have $f''(x_i) \in \{h_1, \dotsc, h_m\}$, only one of which is in $f'(S_i)$, we must have $f''(x_i) = h_j$ as well (in particular, $f''(x_i) = f'(x_i)$). Similarly, $f''(y_{i}) = f'(y_{i}) = z_{j}$. Otherwise, Spoiler wins with $3$ additional pebbles (reusing the pebble on $g$) and $4$ additional rounds, for a total of $5$ rounds. Spoiler now pebbles $(x_{i}, y_{i}) \mapsto (h_{j}, z_{j})$. The pebbled map $(x_i,y_i,g) \mapsto (f'(x_i), f'(y_i), f'(g))$ does not extend to an isomorphism by the choice of $g$, and so Spoiler wins with at most $3$ additional pebbles and $5$ additional rounds.

\item By (c), we may assume that $f'|_{S_{i}}$ is an isomorphism; otherwise, Spoiler wins with $3$ additional pebbles and $5$ additional rounds. In this case, Spoiler pebbles $(g, gxg^{-1}) \mapsto (f'(g), f'(gxg^{-1}))$. Let $f'' : G \to H$ be the bijection that Duplicator selects on the next round. As we pebbled $gxg^{-1} \neq 1$, we have that $f''(S_i)$ and $f'(S_i)$ overlap at $f'(gxg^{-1})$, so we have $f''(S_i) = f'(S_i)$ by \Lem{LemmaSimpleOverlap} (otherwise, by similar argument as in part (c), Spoiler wins with $2$ additional pebbles and $3$ additional rounds).
 
As in the previous part, by (b), we have that $f''(x_{i}) = f'(x_{i}) = h_{j}$ and $f''(y_{i}) = f'(y_{i}) = z_{j}$. Otherwise, Spoiler may win. The strategy in (b) requires Spoiler to use $3$ additional pebbles, and $4$ additional rounds. As the first round of (b) is the second round here in part (d), this yields a total of $5$ rounds. Spoiler now pebbles $(x_{i}, y_{i}) \mapsto (h_{j}, z_{j})$.
Now, as $x \in \langle x_i, y_i \rangle$ there is a word $w$ such that $x = w(x_i, y_i)$. Suppose $\overline{f}$ is a map extending our pebbled map $(x_i, y_i, g, gxg^{-1}) \mapsto (h_j, z_j, f'(g), f'(gxg^{-1}))$. If $\overline{f}(g^{-1}) \neq \overline{f}(g)^{-1}$, then $\overline{f}$ is not an isomorphism. If $\overline{f}(x) \neq w(\overline{f}(x_i), \overline{f}(y_i))$, then $\overline{f}$ is also not an isomorphism. But if $\overline{f}(g^{-1}) = \overline{f}(g)^{-1}$ and $\overline{f}(x) = w(\overline{f}(x_i), \overline{f}(y_i)) = w(h_j, z_j) = f'(x)$, then we have $\overline{f}(gxg^{-1}) \neq \overline{f}(g) \overline{f}(x) \overline{f}(g)^{-1}$ by assumption, so again $\overline{f}$ is not an isomorphism. Thus, there is no isomorphism $\overline{f}$ extending our pebbled map, and Spoiler wins with at most $4$ additional pebbles and $5$ additional rounds. \qedhere
\end{enumerate}
\end{proof}

\Lem{SemisimpleFactors} provides enough to establish that Spoiler can force Duplicator to select at each round a bijection that restricts to an isomorphism on the socles.

\begin{prop} \label{SemisimpleSocleIso}
Let $G$ and $H$ be semisimple groups with isomorphic socles, with $\Fac(\Soc(G)) = \{S_1, \dotsc, S_m\}$ and $S_i = \langle x_i, y_i \rangle$. Let $f : G \to H$ be the bijection that Duplicator selects at some round, and suppose that (i) for all $i$, $f(S_i) \cong S_i$ (though $f|_{S_i}$ need not be an isomorphism) and $f(S_i) \in \Fac(\Soc(H))$, (ii) for every $s \in G$, $\wt(s) = \wt(f(s))$, and
(iii) for all $i$, $f(S_i) = \langle f(x_i), f(y_i) \rangle$. Now suppose that Spoiler pebbles $(x_{1} \cdots x_{m}, y_{1} \cdots y_{m}) \mapsto (f(x_{1} \cdots x_{m}), f(y_{1} \cdots y_{m}))$. 

Let $f' \colon G \to H$ be the bijection that Duplicator selects at any subsequent round in which the pebbles used above have not moved. Then $f'|_{\Soc(G)}\colon \Soc(G) \to \Soc(H)$ must be an isomorphism, or Spoiler can win in $5$ more rounds using at most $5$ more pebbles in the WL$_{II}^{2}$ pebble game.
\end{prop}

\begin{proof}
By Lemma~\ref{LemmaSemisimpleWeight} $f$ must map $\Soc(G)$ bijectively to $\Soc(H)$ and must preserve weights (otherwise, Spoiler wins with $4$ pebbles and $4$ rounds). Suppose that $f|_{\Soc(G)} \colon \Soc(G) \to \Soc(H)$ is not an isomorphism. The only way this is possible is if $f|_{\Soc(G)}$ is not a homomorphism. So there exist $s,s' \in \Soc(G)$ such that $f(s)f(s') \neq f(ss')$.  We claim that this implies that there are $t_i \in S_i$ for all $i=1,\dotsc,m$ such that $f(t_1 t_2 \dotsb t_m) \neq f(t_1) f(t_2) \dotsb f(t_m)$. 

For suppose otherwise. Write $s = s_1 s_2 \dotsb s_m$ and $s' = s_1' s_2' \dotsb s_m'$ with $s_i, s_i' \in S_i$ for all $i$. Then the failure to be a homomorphism is rewritten as:
\[
f(s_1 s_2 \dotsb s_m)f(s_1' s_2' \dotsb s_m') \neq f(s_1 \dotsb s_m s_1' \dotsb s_m') = f(s_1 s_1' s_2 s_2' \dotsb s_m s_m').
\]
Then by our supposition, the LHS of the preceding displayed equation is:
\[
f(s_1) f(s_2) \dotsb f(s_m) f(s_1') f(s_2') \dotsb f(s_m'),
\]
while the RHS is $f(s_1 s_1') f(s_2 s_2') \dotsb f(s_m s_m')$. 
 
 But, by \Lem{SemisimpleFactors}(c), $f$ is an isomorphism on each simple normal subgroup of $\Soc(G)$ (otherwise, Spoiler wins with $4$ additional pebbles and $5$ additional rounds), so the latter is 
 \[
 f(s_1) f(s_1') f(s_2) f(s_2') \dotsb f(s_m) f(s_m'). 
 \]
Furthermore, since the simple normal subgroups of $\Soc(G)$ commute with one another, their images must commute in $\Soc(H)$ (otherwise Spoiler can pebble two elements that commute in $\Soc(G)$ but whose images don't commute in $\Soc(H)$ and win), the LHS is equal to the same, a contradiction. Thus we have some $t_i \in S_i$ for $i=1,\dotsc,m$ such that $f(t_1 t_2 \dotsb t_m) \neq f(t_1) f(t_2) \dotsb f(t_m)$.
 
Suppose $t = t_1 \dotsb t_m$ has weight $w$, and without loss of generality, re-index the $S_i$ and $t_i$ so that $t \in S_1 \times \dotsb \times S_w$ with $t = t_1 t_2 \dotsb t_w$.  We still have $f(t_1 \dotsb t_w) \neq f(t_1) \dotsb f(t_w)$. If $f(t)$ is not in $f(S_1) \times f(S_{i_2}) \times \dotsb \times f(S_{i_w})$ for any $(w-1)$ tuple of distinct indices $i_2, i_3, \dotsc, i_w$, then Spoiler pebbles $t$ and $t_1$. On the next round, Duplicator picks another $f'$ with $f'(t_1) = f(t_1) \in S_1$ and $f'(t) = f(t) \notin f(S_1) \times f(S_{i_2}) \times \dotsb \times f(S_{i_w})$ for any $(w-1)$ tuple of distinct indices. So $f'(t)$ is not the product of something in $f(S_1)$ with any element of weight $w-1$. But by Lemma~\ref{LemmaSemisimpleWeight}, $f(t_2 t_3 \dotsb t_w)$ must have weight $w-1$. Otherwise, Spoiler wins with $3$ additional pebbles beyond those on $(t, t_1)$ and $4$ additional rounds, bringing our total to $5$ additional pebbles and $5$ additional rounds. While \Lem{LemmaSemisimpleWeight} prescribes $4$ additional pebbles, we may reuse the pebble on $t_{1}$. 

Furthermore, we have $f(S_1) = f'(S_1)$ by \Lem{LemmaSimpleOverlap}, or Spoiler can win using 3 additional pebbles and 3 additional rounds, by reusing the pebble on $t$. So we have $f'(t) \neq f'(t_1) f'(t_2 t_3 \dotsb t_w)$. Spoiler may now pebble $t_2 t_3 \dotsb t_w$ and win (3 pebbles, 2 rounds). Thus, $f(t)$ must be the product of something in $f(S_1)$ with some element of weight $w-1$. So there is some $w-1$-tuple of distinct indices $i_2, \dotsc, i_w$ such that $f(t) \in f(S_1) \times f(S_{i_2}) \times \dotsb \times f(S_{i_w})$.

Now repeat the same argument for any $j \in \{1,\dotsc, w\}$ taking the place of $1$. We thus find that for any such $j$, $f(t)$ must be in the direct product of $f(S_j)$ and $w-1$-many other $f(S_i)$'s. Thus $f(t)$ must be in $f(S_1) \times f(S_2) \times \dotsb \times f(S_w)$, since it has weight exactly $w$. So we have $f(t) = f(t_1') f(t_2') \dotsb f(t_w')$ for some $t_i' \in S_i$. But we also have $f(t) \neq f(t_1) f(t_2) \dotsb f(t_w)$. Thus, there must be some $i$ such that $t_i \neq t_i'$. Without loss of generality (by re-indexing if needed), for simplicity of notation we may suppose that $i=1$. Spoiler pebbles $(t,t_1) \mapsto (f(t), f(t_1))$. On the next round, Duplicator selects a new bijection $f'$ with $f'(t) = f(t)$ and $f'(t_1) = f(t_1)$. Since $S_1$ is the unique simple normal subgroup of $\Soc(G)$ containing $t_1$, and similarly $f(S_1)$ is the unique simple  normal subgroup of $\Soc(H)$ containing $f(t_1)$, by \Lem{LemmaSimpleOverlap} we have $f'(S_1) = f(S_1)$. Otherwise, Spoiler wins with $3$ additional pebbles and $3$ additional rounds, for a total of $5$ pebbles and $4$ rounds. While \Lem{LemmaSimpleOverlap} prescribes $4$ additional pebbles, we may reuse the pebble on $t$. 

Finally, we claim that $f'(t_1 t_2 \dotsb t_w) \neq f'(t_1) f'(t_2 t_3 \dotsb t_w)$. Suppose otherwise. Then by Lemma~\ref{LemmaSemisimpleWeight}, $\wt(f'(t)) = \wt(f(t)) = w$, and $\wt(f'(t_2 t_3 \dotsb t_w)) = w-1$ (or Spoiler wins with $3$ additional pebbles and $4$ additional rounds; for the former, Spoiler can re-use the pebble on $t_1$, and for the latter Spoiler can re-use the pebble on $t$). Now, if $f'(t_2 t_3 \dotsb t_w)$ has a non-identity component in the $f(S_1)$ factor of $\Soc(H)$, then there must be some other $i \in \{2, \dotsc, w\}$ such that $f'(t_2 t_3 \dotsb t_w)$ has trivial projection onto $f(S_i)$. Note that $f'(t_1) = f(t_1) \in f(S_1)$ thus also has trivial projection onto $f(S_i)$. But then this would contradict the fact that $\wt(f'(t)) = w$. So the projection of $f'(t_2 t_3 \dotsb t_w)$ onto $f(S_1)$ (as a quotient of $\Soc(H)$) must be trivial. Thus, the component of $f'(t) = f(t)$ in $f(S_1)$ is $f(t_1')$, but the component of $f'(t_1) f'(t_2 \dotsb t_w)$ in $f(S_1)$ is precisely $f'(t_1) = f(t_1) \neq f(t_1')$. Thus $f'(t) \neq f'(t_1) f'(t_2 \dotsb t_w)$ as claimed. As $t,t_1$ are already pebbled, Spoiler now pebbles $t_2 t_3 \dotsb t_w$, and wins ($3$ pebbles, $2$ rounds). The worst strategy Spoiler needed to use was 5 pebbles and $5$ rounds (coming from our application of \Lem{LemmaSemisimpleWeight} in the fourth paragraph).
\end{proof}

\begin{rem}
Brachter \& Schweitzer \cite[Lemma~5.22]{BrachterSchweitzerWLLibrary} previously showed that (1-ary) Weisfeiler--Leman can decide whether two groups have isomorphic socles. However, their results did not solve the search problem; that is, they did not show Duplicator must select bijections that restrict to an isomorphism on the socle even in the case for semisimple groups. This contrasts with \Prop{SemisimpleSocleIso}, where we show that 2-ary WL effectively solves the search problem. This is an important ingredient in our proof that the $(7, 7)$-WL$_{II}^{2}$ pebble game solves isomorphism for semisimple groups.
\end{rem}

We obtain as a corollary of \Lem{SemisimpleFactors} and \Lem{SemisimpleSocleIso} that if $G$ and $H$ are semisimple, then Duplicator must select bijections that restrict to isomorphisms of $\pker(G)$ and $\pker(H)$.

\begin{cor} \label{CorPKerIso}
Let $G$ and $H$ be semisimple groups of order $n$. Let $\Fac(\Soc(G)) := \{ S_{1}, \ldots, S_{m}\}$, and suppose that $S_{i} = \langle x_{i}, y_{i} \rangle$. Let $x := x_{1} \cdots x_{m}$ and $y := y_{1} \cdots y_{m}$. Let $f : G \to H$ be the bijection that Duplicator selects. Spoiler begins by pebbling $(x, y) \mapsto (f(x), f(y))$. Let $f' : G \to H$ be the bijection that Duplicator selects at any subsequent round in which the preceding pebbles have not moved. 
If $f'|_{\pker(G)} : \pker(G) \to \pker(H)$ is not an isomorphism, then Spoiler can win with $5$ additional pebbles and $5$ additional rounds in the WL$_{II}^{2}$ pebble game.
\end{cor}

\begin{proof}
We have the following cases.
\begin{itemize}
\item \textbf{Case 1:} Suppose that $f'(\pker(G)) \neq \pker(H)$. Then without loss of generality, there exists $g \in \pker(G)$ such that $f'(g) \not \in \pker(H)$. By Proposition~\ref{PropSocleSemisimple} we may assume that $f'(S_i) \in \Fac(\Soc(H))$ for all $S_i \in \Fac(\Soc(G))$, or Spoiler can win with $2$ additional pebbles beyond those on $(x, y)$, which we reuse, and 3 additional rounds.

Let $i \in [m]$ such that $f'(g) \cdot f'(S_{i}) \cdot f'(g)^{-1} \neq f'(S_{i})$. Spoiler pebbles $(g, x_{i}) \mapsto (f'(g), f'(x_{i}))$. Let $f'' : G \to H$ be the bijection that Duplicator selects at the next round. As $x_{i} \mapsto f'(x_{i})$ is pebbled, we have that $f''(S_{i}) = f'(S_{i})$; otherwise, by \Lem{LemmaSimpleOverlap}, Spoiler wins with $2$ additional pebbles from the start of this corollary (re-using the pebbles on two of $x,y,g$) and $3$ additional rounds. By \Lem{SemisimpleFactors}(b), we may assume that $f''(y_i) = f'(y_i)$. Otherwise, Spoiler wins with $2$ additional pebbles (reusing the pebbles on $g$ and $x_i$, bringing us to a total of $4$ pebbles) and $4$ additional rounds (for a total of $5$ rounds). Spoiler now pebbles $y_{i} \mapsto f''(y_{i}) = f'(y_{i})$. As $g \in \pker(G)$, we have that $g \cdot S_{i} \cdot g^{-1} = S_{i}$, while $f'(g) \cdot f'(S_{i}) \cdot f'(g)^{-1} \neq f'(S_{i})$. Since the pebbles on $x_i, y_i$ fix their images in $f'(S_i)$, and we have also pebbled $g$, there is no marked isomorphism that extends the pebbling, and thus Spoiler wins, with a total of at most $4$ additional pebbles and $5$ additional rounds.

\item \textbf{Case 2:} Suppose now that $f'({\pker(G)}) = \pker(H)$, but that $f'|_{\pker(G)}$ is not an isomorphism. As $G$ and $H$ are semisimple, so are $\pker(G)$ and $\pker(H)$. Thus, if $f'|_{\pker(G)}$ fails to be an isomorphism, then by \Lem{CharacterizeSemisimple}, there exists $g \in \pker(G)$ and $s \in \Soc(G)$ such that $f'(gsg^{-1}) \neq f'(g) f'(s) f'(g)^{-1}$. By Prop.~\ref{SemisimpleSocleIso}, $f'$ must be an isomorphism on the socle (or Spoiler wins with $5$ additional pebbles and $5$ additional rounds), so if we write $s = s_1 s_2 \dotsc s_m$ with each $s_i \in S_i$, then we must have some $i \in [m]$ such that $f'(g s_i g^{-1}) \neq f'(g) f'(s_i) f'(g)^{-1}$. Since $g \in \pker(G)$, we have $g S_i g^{-1} = S_i$ as well. This case is handled precisely by \Lem{SemisimpleFactors}(d), in which only $4$ additional pebbles and $5$ additional rounds. Thus, for Case 2, we have used at most $5$ additional pebbles and $5$ additional rounds. \qedhere
\end{itemize}
\end{proof}

We now show that if $G$ and $H$ are not permutationally equivalent in their actions on $\Fac(\Soc(G))$ (resp., $\Fac(\Soc(H))$), then Spoiler can win.

\begin{lem} \label{PermutationalIso}
(Same assumptions as \Lem{SemisimpleFactors}.)  Let $G$ and $H$ be semisimple groups with isomorphic socles, with $\Fac(\Soc(G)) = \{S_1, \dotsc, S_m\}$ and $S_i = \langle x_i, y_i \rangle$. Let $f : G \to H$ be the bijection that Duplicator selects at some round, and suppose that (i) for all $i$, $f(S_i) \cong S_i$ (though $f|_{S_i}$ need not be an isomorphism) and $f(S_i) \in \Fac(\Soc(H))$, (ii) for every $s \in G$, $\wt(s) = \wt(f(s))$, and
(iii) for all $i$, $f(S_i) = \langle f(x_i), f(y_i) \rangle$. Now suppose that Spoiler pebbles $(x_{1} \cdots x_{m}, y_{1} \cdots y_{m}) \mapsto (f(x_{1} \cdots x_{m}), f(y_{1} \cdots y_{m}))$. 

Let $f' : G \to H$ be the bijection that Duplicator selects at the next round. Suppose that there exist $g \in G$ and $i \in [m]$ such that $f'(gS_{i}g^{-1}) = f'(S_{j})$, but $f'(g)f'(S_{i})f'(g)^{-1} = f'(S_{k})$ for some $k \neq j$. Then Spoiler can win with $4$ additional pebbles and $4$ additional rounds in the WL$_{II}^{2}$ pebble game.

Furthermore, if $g$ is already pebbled, then Spoiler can win with $3$ additional pebbles and $5$ additional rounds.
\end{lem}

\begin{proof}
We consider the following cases.
\begin{itemize}
\item \textbf{Case 1:} Suppose first that $i = j$. 
In this case, Spoiler pebbles $(g, x_{i}) \mapsto (f'(g), f'(x_{i}))$. Let $f'' : G \to H$ be the bijection that Duplicator selects at the next round. As $x_{i} \mapsto f'(x_{i})$ is pebbled, we have that $f''(S_{i}) = f'(S_{i})$ by Lem.~\ref{LemmaSimpleOverlap}. Otherwise,  Spoiler wins with $3$ additional pebbles, reusing the pebble on $g$, and $3$ additional rounds (bringing our total to $4$ rounds). Spoiler now pebbles $y_{i} \mapsto f''(y_{i})$. The map $(g, x_{i}, y_{i}) \mapsto (f'(g), f'(x_{i}), f''(y_{i}))$ does not extend to a marked isomorphism, as conjugation by $g$ sends $S_i$ to itself, but conjugation by $f'(g)$ does not send $f'(S_i)$ to itself. Thus, in this case, Spoiler wins with at most $4$ additional pebbles and $4$ additional rounds (which is achieved via the application of Lem.~\ref{LemmaSimpleOverlap}).

\item \textbf{Case 2:} Suppose $i \neq j$ but $i=k$. This is symmetric to the preceding case, by swapping the roles of $G$ and $H$. 

\item \textbf{Case 3:} Suppose now that $i, j, k$ are all distinct. By \Lem{SocleDirectProductStronger}, we have that $f'(x_{i}x_{j}) = f'(x_{i})f'(x_{j})$ (or Spoiler wins with $4$ additional pebbles and $5$ additional rounds). Spoiler begins by pebbling $(g, x_{i}x_{j}) \mapsto (f'(g), f'(x_{i}x_{j}))$. Let $f'' : G \to H$ be the bijection that Duplicator selects at the next round. We have the following cases.
\begin{itemize}
\item \textbf{Case 3(a):} Suppose that $f''(S_{i}) = f'(S_{i})$. As $x_{i}x_{j} \mapsto f'(x_{i})f'(x_{j})$ is pebbled, we have necessarily that $f''(x_j) \in f'(S_j)$, and hence that $f''(S_{j}) = f'(S_{j})$ by Lem.~\ref{LemmaSimpleOverlap}. Otherwise, Spoiler wins with $3$ additional pebbles, reusing the pebble on $g$, and $3$ additional rounds (for a total of $4$ rounds). In this case, Spoiler pebbles $(x_{i}, y_{i}) \mapsto (f''(x_{i}), f''(y_{i}))$. As $gS_{i}g^{-1} = S_{j}$, but $f'(g)f''(S_{i})f'(g)^{-1} \neq f''(S_{j})$, the map \\
$(g, x_{i}, y_{i}, x_{i}x_{j}) \mapsto (f'(g), f''(x_{i}), f''(y_{i}), f'(x_{i}x_{j}))$ does not extend to an isomorphism. So Spoiler wins with at most $4$ additional pebbles and $4$ additional rounds.

\item \textbf{Case 3(b):} Suppose that $f''(S_i) \neq f'(S_i)$. As $x_{i}x_{j} \mapsto f'(x_{i})f'(x_{j})$ is pebbled, we have necessarily that $f''(S_{i}) = f'(S_{j})$ and $f''(S_{j}) = f'(S_{i})$; otherwise, by Prop.~\ref{PropSocleSemisimple}, Spoiler wins with $3$ additional pebbles, reusing the pebble on $g$, and $4$ additional rounds. We now have the following sub-cases.
\begin{itemize}
\item \textbf{Case 3(b).i:} Suppose first that $gS_{j}g^{-1} = S_{i}$. By assumption, $gS_{i}g^{-1} = S_{j}$, but $f'(g)f'(S_{i})f'(g)^{-1} = f'(S_{k})$ (where $k \not \in \{i,j\}$). So the conjugation map $\sigma_{g} : a \mapsto gag^{-1}$ swaps $S_{i}$ and $S_{j}$, while the conjugation map $\sigma_{f'(g)}$ does not swap $f'(S_{i})$ and $f'(S_{j})$. It follows that the map $(g, x_{j}, y_{j}, x_{i}x_{j}) \mapsto (f'(g), f''(x_{j}), f''(y_{j}), f'(x_{i}x_{j}))$ does not extend to an isomorphism. Spoiler now pebbles $(x_{j}, y_{j}) \mapsto (f''(x_{j}), f''(y_{j}))$.  Thus, Spoiler wins with at most $4$ additional pebbles and $2$ additional rounds. \\

\item \textbf{Case 3(b).ii:} Suppose now that $gS_{j}g^{-1} \neq S_{i}$. By assumption, as $gS_{i}g^{-1} = S_{j}$, we have that $g^{-1}S_{j}g = S_{i}$. Now recall that $g \mapsto f'(g)$ is pebbled. As the conjugation map $x \mapsto f'(g)^{-1} \cdot x \cdot f'(g)$ is a bijection and $f'(g)f'(S_{i})f'(g)^{-1} = f'(S_{k})$ (where again, $k \not \in \{i, j\}$), we have that $f'(g)^{-1}f'(S_{j})f'(g) \neq f'(S_{i})$.  \\

Suppose first that $f'(g)^{-1}f'(S_{j})f'(g) = f'(S_{j})$. Spoiler now pebbles $(x_{i}, y_{i}) \mapsto (f''(x_{i}), f''(y_{i}))$. As $gS_{i}g^{-1} \neq S_{i}$ and $f''(S_{i}) = f'(S_{j})$, the map $(g, x_{i}, y_{i}, x_{i}x_{j}) \mapsto (f'(g), f''(x_{i}), f''(y_{i}), f'(x_{i}x_{j}))$ does not extend to an isomorphism. So Spoiler wins  with at most $4$ pebbles and $2$ rounds. \\

Suppose now that $f'(g)^{-1}f'(S_{j})f'(g) \neq f'(S_{j})$. By the first paragraph of Case 3(b).ii, we have that $f'(g)^{-1}f'(S_{j})f'(g) \neq f'(S_{i})$. As $g^{-1}S_{j}g = S_{i}$, we have that $f''(g^{-1}x_{j}g) \in f''(S_{i})$. So in particular, $f''(g^{-1}x_{j}g) \neq f'(g)^{-1}f''(x_{j})f'(g)$. Spoiler now pebbles $(x_{j}, g^{-1}x_{j}g) \mapsto (f''(x_{j}), f''(g^{-1}x_{j}g))$. So Spoiler wins with at most $4$ pebbles and $2$ rounds.

\end{itemize} 
\end{itemize}
\end{itemize}

In total, Spoiler used at most $4$ additional pebbles and $5$ additional rounds.
\end{proof}

\begin{thm} \label{SemisimpleIsomorphism}
Let $G$ be a semisimple group and $H$ an arbitrary group of order $n$, not isomorphic to $G$. Then Spoiler has a winning strategy in the $(7,7)$-WL$_{II}^{2}$ pebble game.
\end{thm}

\begin{proof}
If $H$ is not semisimple, then by \Prop{IdentifySemisimple}, Spoiler wins with $3$ pebbles and $2$ rounds. So we now suppose $H$ is semisimple.

Let $\Fac(\Soc(G)) = \{S_1, \dotsc, S_k\}$, and let $x_i, y_i$ be generators of $S_i$ for each $i \in [k]$. Let $f_0$ be the bijection chosen by Duplicator. Spoiler pebbles $(x_1 x_2 \dotsb x_k, y_1 y_2, \dotsc, y_k) \mapsto (f_0(x_1 \dotsb x_k), f_0(y_1 \dotsb y_k))$. On subsequent rounds, we thus have satisfied the hypotheses of \Lem{SemisimpleFactors} and \Prop{SemisimpleSocleIso}. Spoiler will never move these pebbles, 
and thus all subsequent bijections chosen by Duplicator must restrict to isomorphisms on the socle (or Spoiler wins with at most $5$ additional pebbles and $5$ additional rounds. 
Let $f$ be the bijection chosen by Duplicator at the next round. In particular, we may assume that $f$ restricts to an isomorphism on the socle, or Spoiler wins, having used at most $7$ pebbles and $6$ rounds by \Prop{SemisimpleSocleIso}.

By \Lem{CharacterizeSemisimple}, we have that $G \cong H$ if and only if there is an isomorphism $\mu\colon \Soc(G) \to \Soc(H)$ that induces a permutational isomorphism $\mu^*\colon G^* \to H^*$. Thus, since $G \not\cong H$, there must be some $g \in G$ and $s \in \Soc(G)$ such that $f(gsg^{-1}) \neq f(g)f(s)f(g)^{-1}$. Write $s = s_1 \dotsb s_k$ with each $s_i \in S_i$ (not necessarily nontrivial). We claim that there exists some $i$ such that $f(gs_i g^{-1}) \neq f(g) f(s_i) f(g)^{-1}$. For suppose not, then we have
\begin{eqnarray*}
f(gsg^{-1}) & = & f(gs_1 g^{-1} gs_2 g^{-1} \dotsb g s_k g^{-1}) \\
& = &  f(gs_1g^{-1})f(gs_2 g^{-1}) \dotsb f(gs_k g^{-1}) \qquad \text{(by \Prop{SemisimpleSocleIso})}\\
& = & f(g) f(s_1) f(g)^{-1} f(g) f(s_2) f(g)^{-1} \dotsb f(g) f(s_k) f(g)^{-1} \\
& = & f(g) f(s_1 \dotsb s_k) f(g)^{-1} = f(g) f(s) f(g)^{-1},
\end{eqnarray*}
a contradiction. For simplicity of notation, without loss of generality we may assume $i=1$, so we now have $f(gs_1 g^{-1}) \neq f(g) f(s_1) f(g)^{-1}$.

We break the argument into cases:

\begin{enumerate}
\item If $gs_1 g^{-1} \in S_{1}$, then we have $gS_1 g^{-1} = S_1$ (any two distinct simple normal factors of the socle intersect trivially), we have by \Lem{SemisimpleFactors}(d) that Spoiler can win with at most $4$ additional pebbles (for a total of $6$ pebbles) and $5$ additional rounds (for a total of $6$ rounds). 

\item If $gs_1 g^{-1} \in S_j$ for $j \neq 1$ and $f(g)f(s_{1})f(g)^{-1} \notin f(S_j)$, we have by \Lem{PermutationalIso} that Spoiler can win with at most $4$ additional pebbles (for a total of $6$ pebbles) and $5$ additional rounds (for a total of $6$ rounds).

\item Suppose now that $gs_1 g^{-1} \in S_j$ for some $j \neq 1$ and $f(g)f(s_{1})f(g)^{-1} \in f(S_j)$. Spoiler begins by pebbling $(g, gs_{1}g^{-1}) \mapsto (f(g), f(gs_{1}g^{-1}))$. Let $f' : G \to H$ be the bijection that Duplicator selects at the next round. As $gs_{1}g^{-1} \in S_{j}$ is pebbled, we have that $f'(S_{j}) = f(S_{j})$ by Lem.~\ref{LemmaSimpleOverlap} (or Spoiler wins with $2$ additional pebbles, by re-using $2$ of the $4$ already placed pebbles, except for the one on $gs_1g^{-1}$, and $3$ additional rounds). Now by assumption, $gS_{1}g^{-1} = S_{j}$ and $f(g)f(S_{1})f(g)^{-1} = f(S_{j})$. So as $g \mapsto f(g)$ is pebbled, we claim that we may assume $f'(S_{1}) = f(S_{1})$. For suppose not; then we have $g^{-1} S_j g = S_1$ but $f'(g)^{-1} f'(S_j) f'(g) = f(g)^{-1} f(S_j) f(g) = f(S_1) \neq f'(S_1)$. But then Spoiler can with win with $3$ additional pebbles (for a total of $7$ pebbles) and $5$ additional rounds (for a total of $7$ rounds) by \Lem{PermutationalIso}. Thus we have $f'(S_1) = f(S_1)$.

In particular, we have that $f'(x_{1}) = f(x_{1})$ and $f'(y_{1}) = f(y_{1})$, by the same argument as in the proof of \Lem{SemisimpleFactors}(c) (or Spoiler wins with $3$ additional pebbles and $5$ additional rounds, bringing us to a total of $7$ pebbles and $7$ rounds). As $S_{1} = \langle x_{1}, y_{1} \rangle$, we have by \Lem{SemisimpleFactors}(c) that $f'(s_{1}) = f(s_{1})$ (or Spoiler wins with $3$ additional pebbles and $5$ additional rounds, bringing us to a total of $7$ pebbles and $7$ rounds). 
Spoiler now pebbles $(x_{1}, y_{1}) \mapsto (f'(x_{1}), f'(y_{1}))$. As the pebbled map $(g, x_{1}, y_{1}, gs_{1}g^{-1}) \mapsto (f(g),  f'(x_{1}), f'(y_{1}), f'(gs_{1}g^{-1}))$ does not extend to an isomorphism, Spoiler wins using $6$ pebbles and $3$ rounds. Thus, Spoiler used at most $7$ pebbles and $7$ rounds. 
\end{enumerate}

In the worst case (Case 3), we use at most $7$ pebbles and $7$ rounds.
\end{proof}

\section{Conclusion}

We exhibited a novel Weisfeiler--Leman algorithm that provides an algorithmic characterization of the second Ehrenfeucht--Fra\"iss\'e game in Hella's \cite{Hella1989, Hella1993} hierarchy. We also showed that this Ehrenfeucht--Fra\"iss\'e game can identify groups without Abelian normal subgroups using $O(1)$ pebbles and $O(1)$ rounds. In particular, within the first few rounds, Spoiler can force Duplicator to select an isomorphism at each subsequent round. This effectively solves the search problem in the pebble game characterization. 

Our work leaves several directions for further research.

\begin{question}
Can the constant-dimensional $2$-ary Wesifeiler--Leman algorithm be implemented in time $n^{o(\log n)}$?
\end{question}

\begin{question}
What is the (1-ary) Weisfeiler--Leman dimension of groups without Abelian normal subgroups?
\end{question}

\begin{question}
Show that the second Ehrenfeucht--Fra\"iss\'e game in Hella's hierarchy can identify coprime extensions of the form $H \ltimes N$ with both $H,N$ Abelian (the analogue of \cite{QST11}). More generally, an analogue of Babai--Qiao \cite{BQ} would be to show that when $|H|,|N|$ are coprime and $N$ is Abelian, that Spoiler can distinguish $H \ltimes N$ from any non-isomorphic group using a constant number of pebbles that is no more than that which is required to identify $H$ (or the maximum of that of $H$ and a constant independent of $N,H$). 
\end{question}

\begin{question}
Let $p > 2$ be prime, and let $G$ be a $p$-group with bounded genus. Show that in the second Ehrenfeucht--Fra\"iss\'e game in Hella's hierarchy, Spoiler has a winning strategy using a constant number of pebbles. This is a descriptive complexity analogue of  \cite{BMWGenus2, IvanyosQ19}. It would even be of interest to start with the case where $G$ has bounded genus over a field extension $K/\mathbb{F}_{p}$ of bounded degree.
\end{question}

In the setting of groups, Hella's hierarchy collapses to some $q \leq 3$, since 3-ary WL can identify all ternary relational structures, including groups. It remains open to determine whether this hierarchy collapses further to either $q = 1$ or $q = 2$. Even if it does not collapse, it would also be of interest to determine whether the $1$-ary and $2$-ary games are equivalent. Algorithmically, this is equivalent to determining whether $1$-ary and $2$-ary WL have the same distinguishing power.

\begin{question}
Does there exist an infinite family of non-isomorphic pairs of groups $\{ (G_{n}, H_{n})\}$ for which Spoiler requires $\omega(1)$ pebbles to distinguish $G_{n}$ from $H_{n}$? We ask this question for the Ehrenfeucht--Fra\"iss\'e games at both the first and second levels of Hella's hierarchy. 
\end{question}

Recall that the game at the first level of Hella's hierarchy is equivalent to Weisfeiler--Leman \cite{CFI, Hella1989, Hella1993}, and so a lower bound against either of these games provides a lower bound against Weisfeiler--Leman, which, as of this writing,  remains open in the setting of groups. More generally, it would also be of interest to investigate Hella's hierarchy on higher arity structures. For a $q$-ary relational structure, the $q$-ary pebble game suffices to decide isomorphism. Are there interesting, natural classes of higher arity structures for which Hella's hierarchy collapses further to some level $q' < q$?

\section*{Acknowledgment}
The authors would like to thank Pascal Schweitzer and the anonymous referees for feedback on earlier versions of some of the results in this paper, as well as discussions more directly relevant to the paper. JAG would like to thank Martin Grohe for helpful correspondence about WL versus ``oblivious WL.'' JAG was partially supported by NSF award DMS-1750319 and NSF CAREER award CCF-2047756 during this work. ML was partially supported by J. Grochow startup funds, NSF award CISE-2047756, and a Summer Research Fellowship through the Department of Computer Science at the University of Colorado Boulder.

\bibliographystyle{alphaurl}
\bibliography{references}

\end{document}